\documentclass[journal, twocolumn,final]{IEEEtran}
\usepackage{amsfonts}
\usepackage{amssymb}
\usepackage{amsmath}
\usepackage{dsfont}
\usepackage{graphicx,subfigure}
\usepackage{bm}
\usepackage{cite}
\usepackage{bigstrut}
\usepackage{float}
\usepackage{balance}
\usepackage{lipsum}
\usepackage{psfrag}
\usepackage{xcolor}
\usepackage{xfrac}
\usepackage{hyperref}
\usepackage[ruled,vlined]{algorithm2e}
\usepackage{tikz}
\usepackage{pgfplots}
\pgfplotsset{compat=1.15}
\usepackage{mathrsfs}
\usetikzlibrary{arrows}
\usetikzlibrary{decorations.pathreplacing}
\usetikzlibrary{fadings}
\newtheorem{theorem}{Theorem}

\newtheorem{lemma}{Lemma}

\newtheorem{proposition}{Proposition}

\usepackage{mathtools}
\SetKwInput{KwInput}{Input}                
\SetKwInput{KwOutput}{Output}

\begin{document}
\IEEEoverridecommandlockouts

\vspace{-1.7cm}\title{ 
	{\footnotesize \color{cyan} N. G. Evgenidis et al., "Hybrid Semantic-Shannon Communications," in IEEE Transactions on Wireless Communications, vol. 23, no. 9, pp. 10926-10940, Sept. 2024, doi: 10.1109/TWC.2024.3376998.\\
	This is the authors’ preprint version and not the final published version. The published version is available by IEEE at the following link:
   \url{https://ieeexplore.ieee.org/document/10477313}\\
}
	Hybrid Semantic-Shannon Communications}
\author{Nikos G. Evgenidis, Nikos A. Mitsiou,~\IEEEmembership{Graduate Student Member,~IEEE}, Sotiris A. Tegos,~\IEEEmembership{Senior Member,~IEEE}, Panagiotis D. Diamantoulakis,~\IEEEmembership{Senior Member,~IEEE}, Panagiotis Sarigiannidis,~\IEEEmembership{Member,~IEEE}, \\ Ioannis Krikidis,~\IEEEmembership{Fellow,~IEEE}, and George K. Karagiannidis,~\IEEEmembership{Fellow,~IEEE}
\thanks{N. G. Evgenidis, N. A. Mitsiou, and P. D. Diamantoulakis are with the Department of Electrical and Computer Engineering, Aristotle University of Thessaloniki, 54124 Thessaloniki, Greece (e-mails: nevgenid@ece.auth.gr, nmitsiou@auth.gr, padiaman@auth.gr).}
\thanks{S. A. Tegos is with the Department of Electrical and Computer Engineering, Aristotle University of Thessaloniki, 54124 Thessaloniki, Greece and also with the Department of Electrical and Computer Engineering, University of Western Macedonia, 50100, Kozani, Greece (e-mail: sotiristegos@ieee.org).}
\thanks{P. Sarigiannidis is with the Department of Electrical and Computer Engineering, University of Western Macedonia, 50100, Kozani, Greece (e-mail: psarigiannidis@uowm.gr).}
\thanks{I. Krikidis is with the Department of Electrical and Computer Engineering, University of Cyprus, 1678 Nicosia, Cyprus (e-mail: krikidis@ucy.ac.cy).}
\thanks{G. K. Karagiannidis is with the Department of Electrical and Computer Engineering, Aristotle University of Thessaloniki, 54124 Thessaloniki, Greece, and also with the Artificial Intelligence \& Cyber Systems Research Center, Lebanese American University (LAU), Lebanon (e-mail: geokarag@auth.gr).}
\thanks{This work was funded from the Smart Networks and Services Joint Undertaking (SNS JU) under European Union's Horizon Europe research and innovation programme (Grant Agreement No. 101096456 - NANCY).}
\thanks{The work of I. Krikidis was funded from the European Research Council (ERC) under the European Union’s Horizon 2020 research and innovation programme (Grant agreement No. 819819).}
\vspace{-.5cm}
}
\maketitle

\begin{abstract}
Semantic communications are considered a promising beyond-Shannon/bit paradigm to reduce network traffic and increase reliability, thus making wireless networks more energy efficient, robust, and sustainable. However, the performance is limited by the efficiency of the semantic transceivers, i.e., the achievable ``similarity'' between the transmitted and received signals. Under strict similarity conditions, semantic transmission may not be applicable and bit communication is mandatory. In this paper, for the first time in the literature, we propose a multi-carrier \textit{Hybrid Semantic-Shannon} communication system where, without loss of generality, the case of text transmission is investigated. To this end, a joint semantic-bit transmission selection and power allocation optimization problem is formulated, aiming to minimize two transmission delay metrics widely used in the literature, subject to strict similarity thresholds. Despite their non-convexity, both problems are decomposed into a convex and a mixed linear integer programming problem by using alternating optimization, both of which can be solved optimally. Furthermore, to improve the performance of the proposed hybrid schemes, a novel association of text sentences to subcarriers is proposed based on the data size of the sentences and the channel gains of the subcarriers. We show that the proposed association is optimal in terms of transmission delay. Numerical simulations verify the effectiveness of the proposed hybrid semantic-bit communication scheme and the derived sentence-to-subcarrier association, and provide useful insights into the design parameters of such systems.
\end{abstract}
\section{Introduction} \label{Sec:Intro}
The sixth generation (6G) of wireless networks is envisioned to support new types of applications, such as digital twins, the metaverse, and Industry 5.0, which pose new challenges to current communication systems \cite{9955312,6G}. Increasing capacity is a promising research direction to address these challenges, but this approach will soon reach its limits. For instance, the ever-increasing demand for more bandwidth leads to an inevitable bottleneck due to severe path loss and the inefficiency of power amplifiers, while more sophisticated hardware on the user side is needed to cope with the increased bandwidth \cite{6G}. Therefore, a different perspective is required to improve the performance of future wireless networks. Considering that in many 6G scenarios the semantics of data is of greater importance than the data itself, a semantic-oriented approach has recently gained attention \cite{9955525}.

Semantic communications can revolutionize the way that communication is achieved, by considering the difference between the
meaning of the transmitted messages and that of the recovered
messages \cite{9955312}. Shannon and Weaver in their seminal work \cite{6773024} identified three levels of communication, level A which tackles the technical problem of how accurately information is transmitted, level B which is the semantic problem and refers to how informative a message is, and level C which is the effectiveness problem. Semantic communications aim to improve the data exchange between two communication parties by incorporating semantic information in the technical problem of level A \cite{6773024}. This is possible by exploiting common knowledge shared a priori between all parties in the form of knowledge bases.
As such, semantic communications can increase reliability, since an error in the bit level does not necessarily implies an error in the semantic level \cite{chafii2023scientific}. 

The utilization of semantic communications is facilitated by recent developments in the field of deep neural networks (DNNs), such as natural language processing (NLP) and image processing. These advances allow the identification of contextual relationships within texts and images, which in turn are used to extract underlying semantic information from the original data \cite{9955312}. Furthermore, the utilization of joint source-channel coding schemes by means of autoencoders is a powerful tool to increase the robustness of semantic communication systems \cite{9398576}. In this case, not only the amount of transmitted data is reduced, but also the transmission is performed in the most effective way, as well. Hence, the use of artificial intelligence techniques provides realistic ways for the creation of context-aware semantic communication systems.     

\subsection{Literature Review}
Semantic communications have recently attracted a lot of research interest. In \cite{9679803,9955312,9955525}, basic formulations and challenges were discussed in order to establish applications and performance metrics of semantic-aware systems. Although there exist some general concepts behind the foundations of semantic communications in information theory \cite{https://doi.org/10.48550/arxiv.2201.01389, 6004632, 9814642}, most of the relevant literature focuses on their applications. The most characteristic of them are related to the extraction of information from images, speech signals, and text. Regarding image transmission, a DNN called DeepJSCC was studied in \cite{8683463}, which, in addition to semantic extraction, uses source-channel coding to improve the performance of the proposed model. Peak-to-average power ratio (PAPR) was also studied for the specific DNN when the semantic communication paradigm is utilized in orthogonal frequency division multiple access (OFDMA) cellular network \cite{10002903}. Moreover, semantic noise, which is a unique characteristic of semantic communications, was considered in \cite{10012843} and a masked autoencoder was proposed as a solution to improve the robustness of the system. Concerning speech, in \cite{9500590, 9450827}, a DNN called DeepSC-S was used and the signal-to-distortion ratio (SDR) performance metric was studied. A different DNN called DeepSC-SR was also proposed for speech recognition, where character-error-rate (CER) and word-error-rate (WER) were investigated as the most appropriate performance metrics in \cite{9685250}. 

\par Semantic image transmission has also attracted attention. In \cite{10158995}, instead of attempting to reconstruct the signals of the original images using two DeepJSCC DNNs, the concept of perceptual understanding was considered by optimizing the MSE and the learned perceptual image patch similarities (LPIPS). 
In the same direction, in \cite{niu2023hybrid}, a hybrid system was considered where a part of the image is compressed and transmitted in the conventional way, while another part is transmitted utilizing DeepJSCC and the receiver combines the two signals to achieve better perceptual similarity. Furthermore, in \cite{9998051}, the problem of quantizing the output of DeepJSCC was studied, while, in \cite{10066513}, a collaborative image transmission was investigated, where different views of the same image are transmitted over a multiple access channel and the receiver combines the signals to maximize the retrieval accuracy of the original image.

\par Text transmission in particular has been the focus of interest in many works due to its frequent use in everyday data. Regarding the utilized performance metrics in text transmission, the most commonly used are the bilingual evaluation understudy (BLEU) score and sentence similarity.
Specifically, in \cite{9747455}, the similarity maximization of the system was considered when energy harvesting is utilized for energy efficiency. In \cite{9832831}, a similar problem without energy harvesting was studied under a recently introduced performance metric called \textit{semantic accuracy}. Furthermore, a DNN called DeepSC was proposed in \cite{9398576} and also used in \cite{9763856} to reduce the amount of the transmitted information while maximizing the mutual information through end-to-end training. In \cite{9252948}, the authors considered a more practical scenario for the constellation arising from DeepSC, by using quantization on the constructed unstructured constellation. Based on the results of \cite{9763856}, an approximation of the semantic similarity curve was studied in \cite{9940394} along with the utilization of DeepSC for NOMA-based systems, while in \cite{10001594} a quality of experience (QoE) maximization problem was analyzed. Moreover, in \cite{9885016}, the design and use of a DeepSC-like DNN were examined for serving two users. In this case, the training loss function was modified to account for both users. Finally, in \cite{9814566}, a different type of DNN was implemented and combined with a forward relay to improve the common base knowledge between the transmitter and receiver, which can generally be different. Again, an autoencoder structure was used for the end-to-end training of the proposed model.
\subsection{Motivation and Contribution}
Semantic communications have the ability to reduce communication traffic by exploiting the inference capabilities of DNNs. However, DNNs are not always able to achieve the desired accuracy, thus the performance of current semantic frameworks is limited by the design and capabilities of the semantic DNNs transceivers, such as DeepSC \cite{9398576}. For instance, architectures such as DeepSC have a limit on the achievable similarity between the input task, e.g., image, text, speech, and the output task, which strongly depends on channel fading, noise, and power allocation \cite{9763856}. As such, when strict similarity between the input and the output is required, semantic communications may not be a viable option and Shannon communications may be mandatory. Pharmaceutical instructions, or connections that link pharmaceutical substances to diseases they are known to treat can be an everyday example of semantic text transmission \cite{pharma}. In such a case, accuracy is vital for health safety \cite{Shone2011-gs}, thus, perfect similarity between the original and reconstructed instructions is required.
This fact indicates that Shannon and semantic communications should cooperate so that each one hinders the disadvantages of the other. The majority of works on semantic communications emphasize on introducing new DNN architectures to achieve better semantic accuracy between the original and reconstructed data, without investigating the communication performance aspect. However, it is equally important to study appropriate power and subcarrier allocation for these semantic communication protocols, which can further improve their performance and enable their coexistence with Shannon communications. To the best of our knowledge, no current work studies the resource allocation of the hybrid semantic-Shannon communication scheme. Motivated by this, in this paper, we investigate the coexistence of Shannon and semantic communications by proposing a multi-carrier hybrid semantic-Shannon scheme. The contributions of this work are listed below:
%
%
%
\begin{itemize}
    \item For the first time in the open literature, we propose a hybrid multi-carrier system that selects between semantic and Shannon communication per subcarrier. The DeepSC model is used for the semantic transmission, but the proposed hybrid scheme is general and can be used to study the coexistence of current networks with any proposed DNN-based semantic architecture in the literature. The bit error rate (BER) metric is also considered to better evaluate the performance of practical systems, which cannot achieve the capacity limit. 
    \item We formulate a general transmission delay optimization problem, studying two related performance metrics the sum of delays and the maximum delay, by jointly selecting Shannon or semantic transmission per subcarrier and optimizing the total power allocation. The selection between the two transmission schemes distinguishes the proposed hybrid scheme from other multi-carrier problems, where only one transmission protocol exists. Due to the peculiarities introduced by the proposed hybrid transceiver, the problem is non-convex. Based on alternating optimization, the problem is separated into a convex optimization and an integer linear programming one, while the closed-form power allocation solution is provided for both studied problems. 

    \item Semantic communications require a specific sentence to be associated with a specific subcarrier. Therefore, we propose a novel association between sentences and subcarriers, which takes into account the size of the sentences and the channel gain of the subcarriers. Also, by utilizing the rearrangement inequality, we prove the optimality of the proposed association in terms of transmission delay. We note that the optimality of the proposed association is proven with respect to the optimal power allocation of the text transmission delay minimization problem. Thus, the association is performed after obtaining the solution of the minimization problem, and a joint power allocation-association approach is redundant.

    \item Simulations validate the optimality of the proposed sentence-to-subcarrier association. Furthermore, the hybrid scheme is shown to outperform the Shannon communication scheme for both performance metrics, while it is shown that semantic is not always preferable to Shannon communication, which further advocates the combination of the two communication paradigms.  
\end{itemize}

\subsection{Structure}\label{sec:structure}
The rest of this paper is organized as follows. Section \ref{sec:SysMod} describes the system model and the two different frameworks that are combined. In Section \ref{sec:Opt}, we formulate the optimization problem of minimizing the transmission delay time and propose a solution method for it. In Section \ref{sec:ParallelSem}, we analyze the different sentence arrangements in the subcarriers and obtain the optimal one. In Section 
\ref{sec:Results}, we present simulation results and discussion on the performance of the hybrid system, while Section \ref{sec:conclusions} concludes the work.

\begin{table*}[ht]
\caption{List of symbols and basic notations}\label{table:notation}
\small
\centering
\resizebox{\textwidth}{!}{
\centering
\begin{tabular}{ ||c|c||c|c||} 
 \hline
 \multicolumn{4}{|c|}{\textbf{Notation list}} \\
 \hline
 \hline
Number of sentences & $P$ & Maximum similarity level of sentences of $l$-th subcarrier  & $M_{l}^{\mathrm{max}}$ \\ 
\hline
 Number of subcarriers & $L$ & Similarity  upper bound of DeepSC & $M_{\mathrm{sat}}$ \\ 
 \hline
 Number of partitions & $N$ & Required SNR level for semantic transmission of sentence $S_{n,l}$ & $\gamma_{n,l}^{\mathrm{th}}$ \\ 
 \hline
 $j$-th sentence & $S_{j}$ & Maximum required SNR level for semantic transmission of $l$-th subcarrier  & $\gamma_{l}^{\mathrm{max}}$ \\ 
 \hline
 $(n,l)$-th sentence & $S_{n,l}$ & Bandwidth per subcarrier & $W$ \\ 
 \hline
 Semantic form of sentence $S_{n,l}$ & $S_{n,l}'$ & Shannon transmission delay of $l$-th subcarrier & $D_{l}$ \\ 
 \hline
 Number of bits in sentence $S_{n,l}$ & $b_{n,l}$ & Semantic transmission delay of $l$-th subcarrier & $\tilde{D}_{l}$ \\ \hline
 Number of characters in sentence $S_{n,l}$  & $u_{n,l}$ & Power allocation of $l$-th subcarrier & $\mathcal{P}_{l}$ \\ 
 \hline
Encoded vector of transmitted symbols in $S_{j}$ & $\mathbf{x_{j}}$ & Total available transmission power & $\mathcal{P}_{\mathrm{tot}}$ \\ 
\hline 
Number of words in sentence $S_{j}$ & $O_{j}$ & Channel coefficient of $l$-th subcarrier & $h_{l}$ \\ 
\hline 
Similarity level between $S_{n,l}$ and $S_{n,l}'$ & $\tilde{M}_{n,l}$ & Shannon transmission binary variable & $a_{l}$ \\ 
\hline 
Required similarity level of sentence $S_{n,l}$ & $M_{n,l}^{\mathrm{th}}$ & Semantic transmission binary variable & $\tilde{a}_{l}$ \\ 
\hline
Assignment of (R) to (L) & (L) $\gets$ (R) & One-to-one equivalency between (L) and (R) & (L) $\longleftrightarrow$ (R) \\
\hline
\end{tabular}
\label{notation}
}
\end{table*}

\section{System Model} \label{sec:SysMod}
Let $S$ be a task to be transmitted between a user and a BS, as shown in Fig. 1. We note that the investigated scenario is a point-to-point transmission, thus the analysis is also valid for uplink transmission from the user to the BS. Without loss of generality, we assume that this task is a text and it consists of $P$ individual sentences. For the data transmission, $L$ subcarriers can be used. As such, the total number of sentences associated with each subcarrier is given as $N = \left \lfloor \frac{P}{L} \right \rfloor$, where $\lfloor \cdot \rfloor$ denotes the floor operator. For simplicity we assume that $P$ is divided exactly, thus $N = \frac{P}{L}$. Notice that this is not equivalent to equal amount of data at each subcarrier, because the data size of every sentence differs from one to another. Also, $S_j$ denotes the $j$-th sentence in order of appearance in the text, where $j \in \{ 1, \cdots, P \}$, while
\begin{figure}[t!]
    \centering
\includegraphics[width=1 \columnwidth]{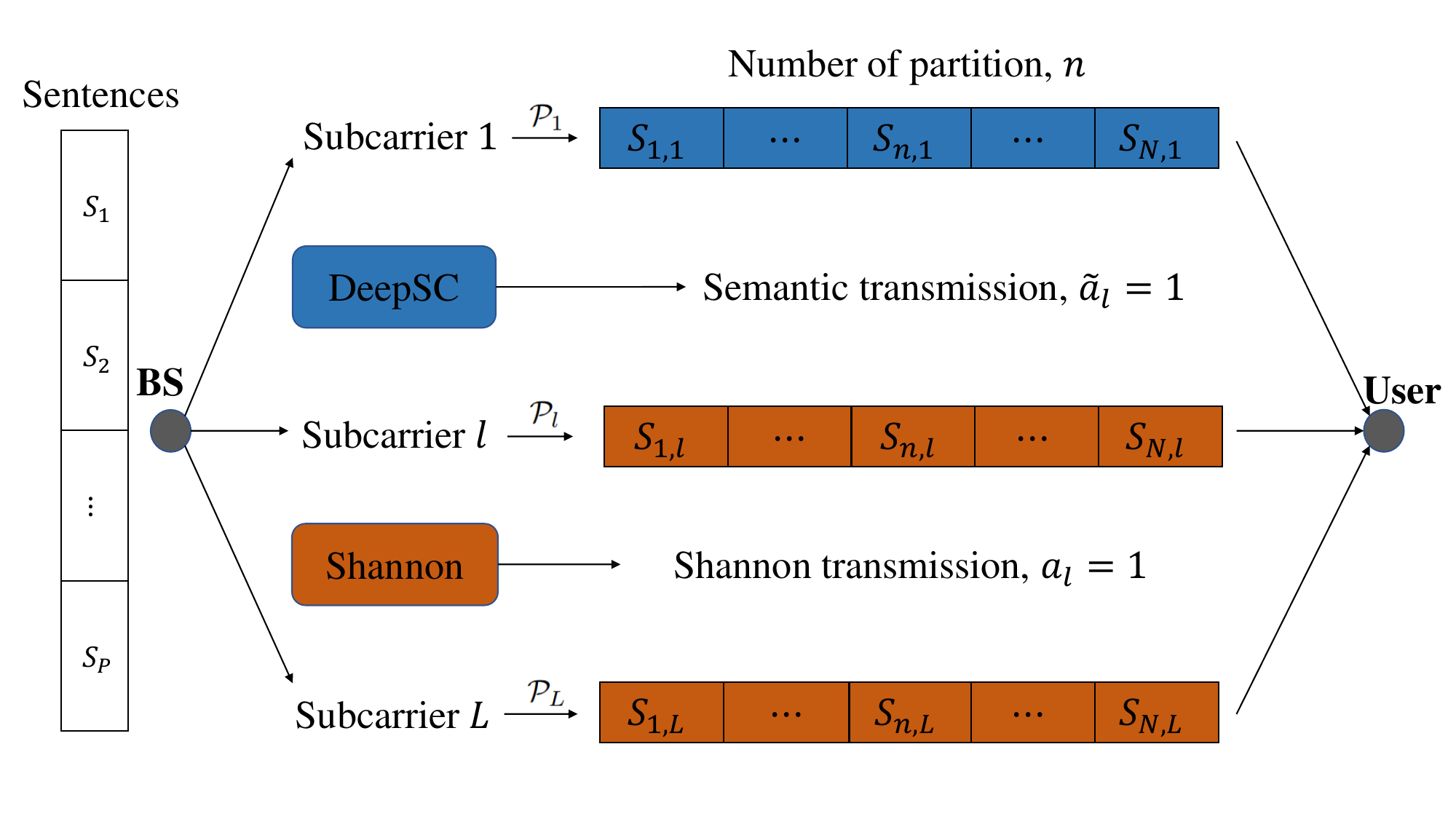}
    \vspace{-7mm}
    \caption{System model.}
    \label{fig:SysModFig}
\end{figure}
$S_{n,l}, n \in \{1,\cdots,N\},l\in\{1,\cdots,L\}$ denotes the $n$-th sentence associated with the $l$-th subcarrier as shown in Fig. \ref{fig:SysModFig}. The two different notations point to the same sentence, i.e., $S_j = S_{n,l}$, and the association between $j$ and the pair $(n,l)$ is given as $j = (n-1)L + l$.
This equivalent representation of sentences is derived from the serial-to-parallel conversion required to assign each sentence to a corresponding subcarrier. For example, if the designed system has $L=16$ subcarriers, then the $25$-th sentence, $S_{25}$, of the text in serial form will be the $2$-nd sentence to be transmitted from the $9$-th subcarrier.
The bandwidth of each subcarrier is denoted as $W$, while $h_{l}$ is the complex channel coefficient, including path loss, between the BS and the user at the $l$-th subcarrier. The user can communicate with the BS by utilizing the principles of either semantic or Shannon communications at each subcarrier as depicted in Fig. \ref{fig:SysModFig}. We note that this choice does not change throughout the coherence time of the channel, but it does change between two different channel instances, while perfect channel state information (CSI) is assumed. The receiver of the proposed system can acknowledge the selected transmission scheme by a single pilot symbol sequence for all subcarriers. For example, two different states of the pilot symbol, i.e., $p_l=+1$ and $p_l=-1$, will symbolize the two different transmission schemes, semantic and Shannon.

\subsection{Shannon Communications} \label{sec:SysModWire}
In Shannon communications, the available sentences are translated to bits, and then transmitted over the wireless medium. As such, the size of sentence $S_{n,l}$ is given as $b_{n,l}$, where the exact value of $b_{n,l}$ depends on the utilized character encoding standard. Assuming the use of the American standard code for information interchange (ASCII), it holds that $b_{n,l} = 8u_{n,l}$, where number $8$ occurs from the fact that $8$ bits are necessary to represent a character, and $u_{n,l}$ is the number of characters of the $S_{n,l}$ sentence.

Shannon-Hartley's theorem states that the maximum capacity between the BS and the user is given by  
\begin{equation}\label{eq:ShannonOriginal}
    C_{l}^{\mathrm{max}} = W \log_2\left( 1+\frac{\mathcal{P}_{l}|h_l|^2}{N_0W} \right),
\end{equation}
where $N_0$ is the power spectral density of additive white Gaussian noise (AWGN), and $\mathcal{P}_{l}$ is the transmission power at each subcarrier. The capacity can be closely approached by capacity-achieving codes such as irregular LDPC codes which are commonly used in practice. Nonetheless, in practical systems, a rate gap between the capacity limit and the achievable maximum data rate may exist, due the bit error rate (BER).  Without loss of generality, we consider the case of an uncoded $M$-QAM constellation, whose BER has been shown in \cite{634685,1177182} to be upper bounded as follows
\begin{equation} \label{eq:BERbound1}
    \mathrm{BER} \leq \frac{1}{5}\exp{\left( - \frac{1.5\mathcal{P}_{l}|h_l|^2}{N_0W} \frac{1}{M-1} \right)},
\end{equation}
where $M$ is the modulation order. 
Consequently, the maximum achievable data rate of an uncoded $M$-QAM scheme, which satisfies a required BER threshold is given by \cite{1177182}
\begin{equation}\label{eq:Shannon}
    C_{l} = W\log_2\left( 1+\frac{\mathcal{P}_{l}|h_l|^2}{N_0W\Gamma} \right),
\end{equation}
where $\Gamma = \sfrac{-\ln({\mathrm{5BER}})}{1.5}$.
We note that $\Gamma \geq 1 $, while for $\Gamma = 1$ \eqref{eq:Shannon} reduces to the Shannon capacity limit.

Therefore, the transmission time delay until all $N$ sentences of the $l$-th subcarrier are transmitted is given as 
\begin{equation} \label{eq:delayWire}
    D_{l} = \frac{U_{l}}{{C_{l}}},
\end{equation}
where $U_{l} = 8\sum_{n=1}^{N} {u_{n,l}}$.
We note that since different sentences have different number of characters, the total size of two partitions associated with two different subcarriers are not equal in general, i.e.,  $\sum_{n=1}^{N} {b_{n,l}} = \sum_{n=1}^{N} {b_{n,l'}}$ does not necessarily hold. 
 
\subsection{Semantic Communications} \label{sec:SysModSem}
As semantic transceiver we adopt the DeepSC model, which was introduced in \cite{9398576} for text transmission. DeepSC utilizes a semantic encoder (decoder) which transforms  sentences to real numbers and vice versa. This is characteristic of the Transformer DNNs and allows easier handling of otherwise difficult to define concepts, like the semantics of a sentence. Moreover, it utilizes a  channel encoder (decoder) which maximizes the mutual information between the transmitter and the receiver. Both semantic and channel encoders (decoders) are jointly trained to maximize the mutual information between the transmitted and the received sentence. Also, since a sentence is mapped to a sequence of continuous real numbers, the output of the channel encoder is an optimal constellation of infinite points, which is equivalent to transmitting the output of the semantic encoder by using discrete time analog transmission (DTAT) \cite{10002903}. As such, the transmission rate of DeepSC is given as 
\begin{equation}
    \tilde{C}_{l} = W.
\end{equation}

For a sentence $S_{n,l}$, we denote $S_{n,l}'$ its semantic equivalent form, i.e., the output of the semantic encoder when the input is sentence $S_{n,l}$. 
Also, to mitigate the effects of AWGN, $S_{n,l}'$ is encoded, through the channel encoder, into the vector $\mathbf{x}_{j} = [x_{1}, \cdots, x_{kO_{j}}]$, where $\mathbf{x}_{j}$ consists of the encoded symbols that need to be transmitted, $O_{j}$ denotes the number of words in the $j$-th sentence and $k$ is the number of outputs of the DNN for each word.  For transmission using the DeepSC model, we denote by $s_{n,l}$ the symbols that need to be transmitted via DTAT. As such, for a fixed number of outputs $k$, the total number of semantic symbols per sentence are $s_{n,l} = kO_{j}$. Consequently, the transmission time delay for semantic transmission is given as 
\begin{equation} \label{eq:delaySem}
    \tilde{D}_{l} = \frac{k\sum_{n=1}^{N} {O_{(n-1)L+l}}}{{\tilde{C}_{l}}},
\end{equation}
due to the relationship between $j$ and $(n,l)$.

\begin{figure}[b]
    \centering
    \includegraphics[width=1\columnwidth]{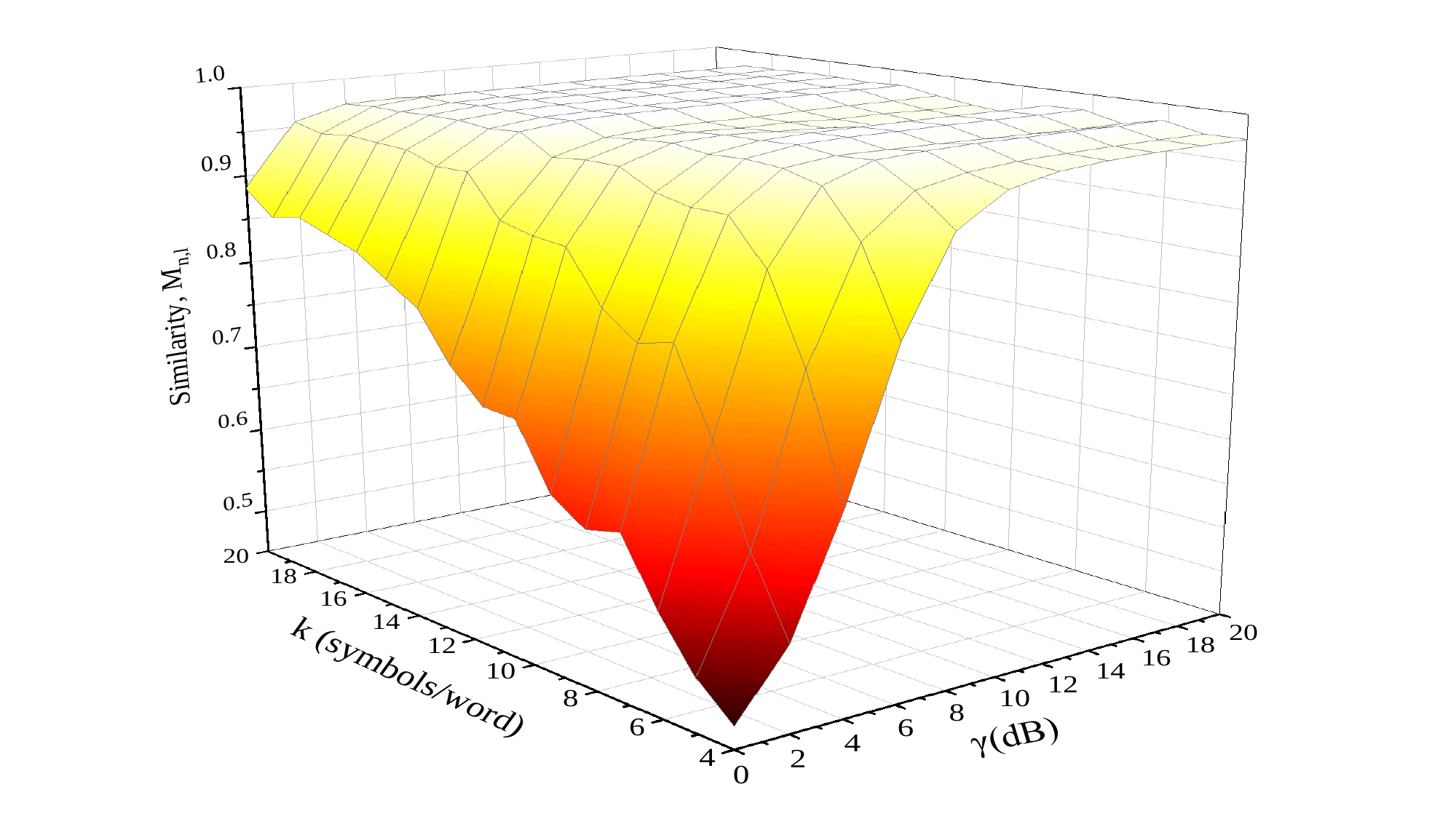}
    \caption{Similarity for varying values of SNR and $k$.}
    \label{fig:SimilarityTotal}
\end{figure}

Semantic communications are affected by both AWGN and fading. However, instead of measuring the absolute value of error, i.e., bits in Shannon communications, semantic communications are interested in measuring the semantic similarity between the transmitted and the received signal. It should be highlighted that although bit errors can occur in digital transmission, coding schemes can be utilized to correct such errors achieving rates up to the capacity $C_l$.
For text transmission, the cosine similarity metric \cite{9763856} is used
\begin{equation}\label{eq:SimDefine}
    M_{n,l} = \frac{\mathbf{B}(S_{n,l}) \mathbf{B}(S'_{n,l})^{T} }{ \|\mathbf{B}(S_{n,l}) \| \| \mathbf{B}(S'_{n,l})^{T} \|},
\end{equation}
where $(\cdot)^{T}$ denotes the transpose operator of a vector, and $\mathbf{B}(\cdot)$ denotes the bidirectional encoder representations from transformers (BERT) for each sentence $S_{n,l}$, which is a vectorized representation of the original sentence after its processing from the DeepSC encoder. Therefore, $M$ provides a measure of similarity between the transmitted and the received sentence. It can be observed in \eqref{eq:SimDefine} that sentence similarity is applied on an entire sentence without restrictions on the sentence's length in contrast to the BLEU score and its dependency on $m$-grams, which allows it to detect semantic relations between words throughout the length of the sentence. For Shannon communications, we assume that no bit errors occur that cannot be corrected by capacity-achieving coding implementation, thus the similarity between the transmitted and the received sentence equals to one, i.e., $M_{n,l} = 1$ \cite{9763856}.  
On the other hand, for semantic communications $\tilde{M}_{n,l} \in (0,1]$, where $\tilde{M}_{n,l}$ symbolizes the similarity between $S_{n,l}'$ and $S_{n,l}$, due to the fact that the semantic meaning of the reconstructed sentence can differ from that of the original one. In fact, DeepSC model cannot achieve any desired value of similarity, but it is upper bounded by a certain similarity level, which is denoted as $M_{\mathrm{sat}}$. The achievable similarity level of DeepSC can be found by training and testing the DNN model as discussed in \cite{9763856}. To achieve the best similarity levels possible and not compromise the performance of the semantic transceiver, DeepSC is trained for individual SNR values as in \cite{9398576}. The resulting similarity levels are shown in Fig. \ref{fig:SimilarityTotal} for varying parameters after training for a range of different SNR values. As it is evident, this similarity level depends on the number of outputs $k$, the received signal-to-noise ratio (SNR), and the training of the DeepSC model. Notice that $k$ is a DNN parameter that must be predetermined for training and, as such, should be chosen to better reflect the desired similarity levels, but also considering that the larger $k$ is, the more symbols for transmission are used.

From \cite{10001594} and Fig. \ref{fig:SimilarityTotal}, it is concluded that for fixed $k$ the function corresponding to the achievable similarity values of DeepSC is an increasing one with respect to the transmit SNR and each similarity threshold $M_{n,l}^{\mathrm{th}}$ is one-to-one mapped to a minimum SNR value $\gamma_{n,l}^{\mathrm{th}}$. Hence, to ensure that the received SNR of all semantic symbols satisfies the minimum threshold, all symbols associated with the same sentence will be transmitted from the same subcarrier, by using the same power allocation policy. Moreover, we note that the practicality of the proposed hybrid scheme can be verified by \cite{9252948}, where a quantized version of the DeepSC model was proposed, which essentially reduces the DeepSC model to a digital communication scheme. By choosing an appropriate number of quantization levels the digital counterpart of DeepSC loses negligible accuracy, while its transmission delay is still defined by the available bandwidth \cite{9252948}. Thus, in our analysis, only the original DeepSC model was presented.

\section{Transmission Delay Metrics Minimization} \label{sec:Opt}

\subsection{General Problem Formulation} \label{sec:OptFor}
Performance metrics associated with data transmission delay are of great importance in current communication systems especially in multi-carrier systems where simultaneous transmission among the subcarriers takes place. The minimization of such metrics, like the maximum delay and the sum of delays, is crucial to several applications of wireless networks, such as federated learning, digital twins, and autonomous vehicles \cite{6G}. Thus, it is of paramount importance to study the impact of the proposed hybrid semantic-Shannon scheme to the transmission delay. We note that text transmission is a specific data communication example, nonetheless the impact of the hybrid semantic-Shannon scheme to the transmission delay can be assumed to be similar for other cases, such as image transmission which is essential to autonomous driving. It should also be highlighted that the presented analysis, by using a different similarity metric, is valid for image transmission. 
Combining \eqref{eq:delayWire} and \eqref{eq:delaySem}, the transmission delay of the $l$-th subcarrier is equal to 
\begin{equation} \label{eq:delayTime}
    \mathcal{D}_{l}= a_{l}D_{l} + \tilde{a}_{l}\tilde{D}_{l},
\end{equation}
where $a_{l}, \tilde{a}_{l} \in \{ 0,1 \}$ are binary decision variables such that $a_{l} + \tilde{a}_{l} = 1, \, \forall l \in \{ 1, \cdots , L \}$, which allow selection between Shannon and semantic communication for each subcarrier individually. Specifically, $a_{l}=1$ denotes that the $l$-th subcarrier utilizes Shannon communication, while $\tilde{a}_{l}=1$ indicates that the $l$-th subcarrier uses the DeepSC model. 
We assume that each sentence $S_{n,l}$ has a similarity threshold of at least $M_{n,l}^{\mathrm{th}}$, which can be considered as a semantic quality of service (QoS) constraint. If this semantic QoS is not guaranteed communication fails.  Therefore, for each sentence $S_{n,l}$ it must be, 
\begin{equation} \label{eq:SimilarityCond}
    a_{l} + \tilde{a}_{l}\tilde{M}_{n,l} \geq M_{n,l}^{\mathrm{th}}, \, 
\forall n \in \{1, \cdots, N \}, \forall l \in \{1, \cdots, L \}.
\end{equation}
In the $l$-th subcarrier, all sentences are bounded by the similarity constraints given by $M_{n,l}^{\mathrm{th}}$. However, since all $N$ sentences per subcarrier are transmitted within the coherence time interval, all $N$ sentences are subject to equal channel fading. Thus, the power allocation of the $l$-th subcarrier must ensure the similarity constraint of all $N$ sentences,  which implies that the maximum similarity of all sentences to be transmitted  per subcarrier must be achievable. As such, when designing the optimal power allocation for the $l$-th subcarrier, the following similarity constraint needs to hold
\begin{equation}\label{eq:maxSim}
    M_{l}^{\mathrm{max}} = \underset{1 \leq n \leq N}{\mathrm{max}} \left\{M_{n,l}^{\mathrm{th}}\right\}
\end{equation} 
for the $l$-th subcarrier to utilize DeepSC. We note that due to the similarity upper bound of DeepSC, when $M_{l}^{\mathrm{max}} > M_{\mathrm{sat}}$, Shannon communications have to be used. For convenience, we define the set of all subcarriers which can use DeepSC as 
\begin{equation}\label{eq:SetPotSem}
    \mathcal{S} = \left\{ l | M_{\mathrm{sat}} \geq M_{l}^{\mathrm{max}} , \, \forall l \right\},
\end{equation}
while  the set of all subcarriers which prefer semantic transmission to the Shannon counterpart is given as
\begin{equation}\label{eq:SetCurSem}
    \mathcal{S}' = \left\{ l | \tilde{a}_{l} = 1 , \, \forall l \right\}.
\end{equation}
Since for each $M_{l}^{\mathrm{max}}$ there exists an one-to-one mapping to a value of $\gamma_{l}^{\mathrm{max}}$, it is straightforward to show that by combining \eqref{eq:SimilarityCond} and \eqref{eq:maxSim}, the following power constraint holds
\begin{equation} \label{eq:SNRcond}
    \mathcal{P}_{l} \geq \gamma_{l}^{\mathrm{max}} c_{l}, \, \forall l \in \mathcal{S}',
\end{equation}
where $c_l = \frac{N_0W}{|h_l|^2}$.

\par Let $\bm{\mathcal{P}} = \left\{ \mathcal{P}_{1}, \mathcal{P}_{2}, \cdots, \mathcal{P}_{L} \right\}$ be the transmission power variables set and in similar fashion $\bm{a} = \left\{ a_{1}, a_{2}, \cdots, a_{L} \right\}$ and $\bm{\tilde{a}} = \left\{ \tilde{a}_{1}, \tilde{a}_{2}, \cdots, \tilde{a}_{L} \right\}$ are the Shannon and semantic binary variables sets, respectively. We also define a delay metric, denoted by the function $f = f(\mathcal{D}_{1},\cdots,D_{l},\cdots,D_{L})$.  Based on the above analysis, we study two delay metrics for the hybrid transmission. We note that studying both metrics further validates the motivation of the paper, i.e., that semantic communications may not always offer improved performance and they should coexist with Shannon communications.

\subsection{Minimization of Sum of Delays}
The sum of delay is a metric widely used in the literature to calculate the average transmission time \cite{8664595,9133107}.
In this case, the objective function that we aim to minimize is the overall transmission time of all $L$ subcarriers, which is given below
\begin{equation} \label{eq:completeBit}     f(\mathcal{D}_{1},...,\mathcal{D}_{l},...,\mathcal{D}_{L}) = \sum_{l=1}^{L} \mathcal{D}_{l}
\end{equation}
and the optimization problem under discussion can be described as 
\begin{equation*}\tag{\textbf{P1}}\label{eq:OptProblem}
    \begin{array}{ll}
    \mathop{\mathrm{min}}\limits_{\bm{\mathcal{P},a,\tilde{a}}}
    &\sum_{l=1}^{L} \mathcal{D}_{l} \\
    \,\,\textbf{s.t.} \quad
    & \mathrm{C}_1: a_{l}\! +\! \tilde{a}_{l}\tilde{M}_{n,l}\! \geq \!M_{n,l}^{\mathrm{th}}, \, \forall (n,l) \\
    & \mathrm{C}_2: a_{l} + \tilde{a}_{l} = 1, \, \forall l \\
    & \mathrm{C}_3: a_{l} ,\tilde{a}_{l} \in \{ 0,1 \}, \, \forall l \\
    & \mathrm{C}_4: \sum_{l=1}^{L} \mathcal{P}_{l} = \mathcal{P}_{\mathrm{tot}}.
    \end{array}
\end{equation*}

It is highlighted that \eqref{eq:OptProblem} is not jointly convex with respect to $\mathcal{P}_{l}$, $a_{l}$, and $\tilde{a_{l}}$, because of the binary variables and the fact that $\sum_{l=1}^{L} \mathcal{D}_{l}$ depends on $\bm{a,\tilde{a}}$. However, we can observe that by the definition of $\sum_{l=1}^{L} \mathcal{D}_{l}$ in \eqref{eq:completeBit}, it is convex in terms of $\mathcal{P}_{l}$ when $a_{l}, \tilde{a}_{l}$ are fixed. Thus, alternating optimization can be used to solve \eqref{eq:OptProblem} by separating the initial problem into two problems, i.e., one convex optimization problem with respect to $\bm{\mathcal{P}}$ and one mixed integer linear problem with respect to $\bm{a,\tilde{a}}$. 

Considering \eqref{eq:SetPotSem}, \eqref{eq:SetCurSem} and \eqref{eq:SNRcond}, the convex optimization problem with respect to ${\mathcal{P}_{l}}$ can be formulated as
\begin{equation*}\tag{\textbf{P2}}\label{eq:OptProblemConvex}
    \begin{array}{ll}
    \mathop{\mathrm{min}}\limits_{\bm{\mathcal{P}}}
    & \sum_{l=1}^{L} \mathcal{D}_{l} \\
    \,\,\text{\textbf{s.t.}}& \mathrm{C}_1: -\mathcal{P}_{l} \leq -\gamma_{l}^{\mathrm{max}} c_{l}, \, \forall l \in \mathcal{S}' \\
    & \mathrm{C}_2:  \sum_{l=1}^{L} \mathcal{P}_{l} = \mathcal{P}_{\mathrm{tot}},
    \end{array}
\end{equation*}
while the mixed integer linear programming problem with respect to $\bm{a,\tilde{a}}$ can be formulated as
\begin{equation*}\tag{\textbf{P3}}\label{eq:OptProblemIntLin}
    \begin{array}{ll}
    \mathop{\mathrm{min}}\limits_{\bm{a,\tilde{a}}}
    & \sum_{l=1}^{L} \mathcal{D}_{l} \\
    \,\,\text{\textbf{s.t.}}
    & \mathrm{C}_1: \tilde{a}_{l}\mathcal{P}_{l} \leq \gamma_{l}^{\mathrm{max}} c_{l}, \, \forall l \in \mathcal{S} \\
    & \mathrm{C}_2: a_{l} + \tilde{a}_{l} = 1, \, \forall l \\
    & \mathrm{C}_3: a_{l} ,\tilde{a}_{l} \in \{ 0,1 \}, \, \forall l \\
    & \mathrm{C}_4: \sum_{l=1}^{L} \mathcal{P}_{l}(a_{l} + \tilde{a}_{l}) = \mathcal{P}_{\mathrm{tot}}.
    \end{array}
\end{equation*}
We note that while in \eqref{eq:OptProblemConvex} there exist inequality power constraints only over $\mathcal{S}'$, \eqref{eq:OptProblemIntLin} should have inequality power constraints over all subcarriers in $\mathcal{S}$. This is necessary  for a subcarrier to be able to successfully select between semantic or Shannon transmission.
Problem \eqref{eq:OptProblemIntLin} can be optimally solved by using standard integer linear programming tools, e.g., Branch and Bound, while a closed-form solution for  \eqref{eq:OptProblemConvex} is given below.
\subsubsection{A Closed-form Power Allocation Solution}\label{sec:ConvexProbSec}

Problem \eqref{eq:OptProblemConvex} can be easily proven to be convex, therefore the Karush-Kuhn-Tucker (KKT) conditions are necessary and sufficient to obtain its optimal value. This leads to the following Proposition \ref{pro:KKTOptimalSol}. 

\begin{proposition} \label{pro:KKTOptimalSol}
The optimal power allocation for the subcarriers utilizing semantic communication is described by
\begin{equation} \label{eq:sempowalloc}
    \mathcal{P}_{l}^* = \gamma_{l}^{\mathrm{max}} c_{l}, \, \forall l \in \mathcal{S}'
\end{equation}
and for the ones utilizing Shannon communication by
\begin{equation}\label{eq:SolveKKT}
    \mathcal{P}_{l}^* = c_{l}\Gamma\exp\left[2 W_0 \left( \sqrt{\frac{\delta}{4}} \right) \right] - 1, \, \forall l \notin \mathcal{S}',
\end{equation}
where $\delta = \frac{\ln{(2)} U_{l}}{\lambda Wc_{l}\Gamma}$ and $W_0(\cdot)$ denotes the principal branch of the Lambert W function, and $\lambda$ is the solution of the following equation 
\begin{equation} \label{eq:lambdaEqSol}
    \sum_{\substack{l=1 \\ l \notin \mathcal{S}'}}^{L} \mathcal{P}_{l}^{*}(\lambda) = \mathcal{P}'_{\mathrm{tot}}. 
\end{equation}  
\end{proposition}

\begin{IEEEproof}
Since $\mathcal{P}_{l},\, \forall l \in \mathcal{S}'$, does not affect the transmission delay of the semantic scheme, but only the similarity, the optimal value of $\mathcal{P}_{l},\, \forall l \in \mathcal{S}'$, is given by \eqref{eq:sempowalloc}.
Then, substituting \eqref{eq:sempowalloc} to $\mathrm{C}_2$ of \eqref{eq:OptProblemConvex}, $\mathrm{C}_2$ is transformed as 
\begin{equation}\label{eq:OnlyWireP}
    \mathrm{C}_2:\sum_{\substack{l=1 \\ l \notin \mathcal{S}'}}^{L} \mathcal{P}_{l} = \mathcal{P}_{\mathrm{tot}} - \sum_{\substack{l=1 \\ l \in \mathcal{S}'}}^{L} \gamma_{l}^{\mathrm{max}} c_{l} = \mathcal{P}'_{\mathrm{tot}},
\end{equation}
where $\mathcal{P}'_{\mathrm{tot}}$ denotes the available power for Shannon transmission. Then, the Lagrangian function of \eqref{eq:OptProblemConvex} can be written as 
\begin{equation}\label{eq:Lagrange}
    \mathcal{L}(\mathcal{P}_{l},\lambda) =  \sum_{\substack{l=1 \\ l \notin \mathcal{S}'}}^{L} \frac{U_{l}}{W\log_2{\left(1 + \frac{\mathcal{P}_{l}}{c_{l}\Gamma} \right)}} - \lambda\left( \sum_{\substack{l=1 \\ l \notin \mathcal{S}'}}^{L} \mathcal{P}_{l} - \mathcal{P}'_{\mathrm{tot}} \right).
\end{equation}
As such, the optimal value of $\mathcal{P}_l, \, \forall l \notin \mathcal{S}'$, can be obtained by solving the following set of equations
\begin{equation}\label{eq:SystemKKT}
    \frac{\partial \mathcal{L}}{\partial \mathcal{P}_{l}} = \frac{\ln{(2)} U_{l}}{Wc_{l}\Gamma \left( 1+\frac{\mathcal{P}_{l}}{c_l\Gamma} \right) \ln^2{ \left( 1+\frac{\mathcal{P}_{l}}{c_l\Gamma} \right )}} - \lambda = 0, \, \forall l \notin \mathcal{S}',
\end{equation}
 Solving with respect to $\mathcal{P}_{l}$ in \eqref{eq:SystemKKT}, the optimal power allocation at the $l$-th subcarrier is given by \eqref{eq:SolveKKT}. Using the total power constraint $\sum_{l=1}^L \mathcal{P}_l = \mathcal{P}_{\mathrm{tot}}$ and substituting \eqref{eq:sempowalloc} and \eqref{eq:SolveKKT}, we get \eqref{eq:lambdaEqSol} from which we can obtain $\lambda$ by means of the bisection method, or using more advanced methods, e.g., the Powell's dog leg method \cite{rabinowitz1970numerical}. Then, substituting $\lambda$ in \eqref{eq:SolveKKT} yields the optimal power allocation.
\end{IEEEproof}

The procedure for jointly deriving the optimal power allocation and semantic-Shannon selection at each subcarrier is summarized in Algorithm \ref{alg:alternateOpt}.

\begin{algorithm}
\caption{Solution of \eqref{eq:OptProblem} via alternating optimization}
Fix number of semantic symbols per word, $k$. \\
Fix number of iterations, $I$. \\
Find $M_{l}^{\mathrm{max}}$ for each subcarrier $l$. \\
Convert between similarity and SNR $M_{l}^{\mathrm{max}} \longleftrightarrow \gamma_{l}^{\mathrm{max}}$. \\
Fix $\bm{a,\tilde{a}}$ to  $a_{l} = 1$ and $\tilde{a}_{l} = 0, \, \forall l$. Keep $\bm{a}^{(1)},\bm{\tilde{a}}^{(1)}$.\\
Solve \eqref{eq:OptProblemConvex} via KKT conditions to attain $\mathcal{P}_{l}, \, \forall l$. \\
\For {$i = 1:I$}{
Solve \eqref{eq:OptProblemIntLin} to attain $\bm{a,\tilde{a}}$. Keep $\bm{a}^{(i+1)},\bm{\tilde{a}}^{(i+1)}$. \\
\If{$\bm{a}^{(i+1)}= \bm{a}^{(i)}$,  $\bm{\tilde{a}}^{(i+1)}= \bm{\tilde{a}}^{(i)}$}{\textbf{break}}
Solve \eqref{eq:OptProblemConvex} to attain $\mathcal{P}_{l}, \, \forall l$.
} 
Keep minimum achieving $\bm{a}^{(i)}, \bm{\tilde{a}}^{(i)}$ and $\mathcal{P}_{l}^{(i)}$.
\label{alg:alternateOpt}
\end{algorithm}

\subsection{Minimization of Maximum of Delays}
The maximum delay metric describes the total transmission duration of the proposed multi-carrier scheme, and it has also been widely studied in the literature \cite{9268953}. 
In this case, the objective function that we aim to minimize is the maximum transmission delay time of all $L$ subcarriers, given as
\begin{equation} \label{eq:maxDelay}  f(\mathcal{D}_{1},...,D_{l},...,D_{L}) = \underset{1 \leq l \leq L}{\max}{ \mathcal{D}_{l}}
\end{equation}
and the formulated optimization problem can be described as
\begin{equation*}\tag{\textbf{P4}}\label{eq:OptProblemMinMax}
    \begin{array}{ll}
    \mathop{\mathrm{min}}\limits_{\bm{\mathcal{P},a,\tilde{a}}}
    &\underset{1 \leq l \leq L}{\max}{ \mathcal{D}_{l}} \\
    \,\,\textbf{s.t.} \quad
    & \mathrm{C}_1: a_{l}\! +\! \tilde{a}_{l}\tilde{M}_{n,l}\! \geq \!M_{n,l}^{\mathrm{th}}, \, \forall (n,l) \\
    & \mathrm{C}_2: a_{l} + \tilde{a}_{l} = 1, \, \forall l \\
    & \mathrm{C}_3: a_{l} ,\tilde{a}_{l} \in \{ 0,1 \}, \, \forall l \\
    & \mathrm{C}_4: \sum_{l=1}^{L} \mathcal{P}_{l} = \mathcal{P}_{\mathrm{tot}}.
    \end{array}
\end{equation*}
As previously, \eqref{eq:OptProblemMinMax} is not convex and, thus, alternating optimization can be used to separately optimize $\mathcal{P}_{l}$ and $a_{l},\tilde{a}_{l}$. For the optimization of $\mathcal{P}_{l}$ we first observe that any subcarrier that utilizes semantic communications will satisfy \eqref{eq:SNRcond} and its delay is not subject to optimization as shown by \eqref{eq:delaySem}. Therefore, semantic communications can be utilized only by subcarriers whose delay is less than that of the subcarriers that utilize Shannon communications, since the latter must have equal delays amongst them. As such, we present a closed-form solution for the optimization of $\mathcal{P}_{l}$ along with a heuristic algorithm for the optimal selection between semantic and Shannon utilization. 
The subcarriers that utilize semantic communications have to allocate such power so that the similarity constraint is satisfied with equality, meaning that whenever holds that $\tilde{a}_{l} = 1$, it also has to hold that 
\begin{equation}
    \label{eq:sempowallocMinMax}
    \mathcal{\tilde{P}}_{l}^* = \gamma_{l}^{\mathrm{max}} c_{l}, \, \forall l \in \mathcal{S}'.
\end{equation}
Then, for the subcarriers that utilize Shannon communications a closed-form solution can be obtained, because the minimization of the maximum delay problem is reduced in an equality problem between all subsequent $D_{l}$.
With this in mind, the following analysis holds for any subcarriers such that $m,l \notin \mathcal{S}'$:
\begin{equation}\label{eq:CondEq}
    \frac{U_{l}}{W\log_{2}\left( 1 + \frac{\mathcal{P}_{l}}{c_{l}\Gamma} \right)} = \frac{U_{m}}{W\log_{2}\left( 1 + \frac{\mathcal{P}_{m}}{c_{m}\Gamma} \right)},
 \end{equation}   
 which yields that
 \begin{equation}\label{eq:PowerCond}   
    P_{m} = c_{m}\Gamma \left( \left( 1 + \frac{\mathcal{P}_{l}}{c_{l}\Gamma} \right)^{\left(\frac{U_{m}}{U_{l}} \right)} - 1 \right).
\end{equation}
From the overall power constraint of the problem, the following condition must hold:
\begin{equation}\label{eq:PowerEq}
     \mathcal{P}_{\mathrm{tot}} - \sum_{\substack{m=1 \\ m \in \mathcal{S}'}}^{L} \gamma_{m}^{\mathrm{max}} c_{m} = \sum_{\substack{m=1 \\ m \notin \mathcal{S}'}}^{L} c_{m}\Gamma\left( \left( 1 + \frac{\mathcal{P}_{l}}{c_{l}\Gamma} \right)^{\left(\frac{U_{m}}{U_{l}} \right)} - 1 \right).
\end{equation}
The last one can be solved in terms of $\mathcal{P}_{l}$ and the rest of the power allocations for the other subcarriers can be found recursively by \eqref{eq:PowerCond}.
Using the aforementioned analysis, we propose a heuristic algorithm to solve \eqref{eq:OptProblemMinMax} aiming to find the optimal selection between semantic and Shannon utilization and their joint power allocation problem.

\begin{algorithm}[!ht]
\caption{Algorithm for selection between semantic and Shannon communication}
Fix number of semantic symbols per word, $k$. \\
Fix number of iterations, $I$. \\
Find $M_{l}^{\mathrm{max}}$ for each subcarrier $l$. \\
Convert between similarity and SNR $M_{l}^{\mathrm{max}} \longleftrightarrow \gamma_{l}^{\mathrm{max}}$. \\
Fix $\bm{a,\tilde{a}}$ to  $a_{l} = 1$ and $\tilde{a}_{l} = 0, \, \forall l$. Keep $\bm{a}^{(1)},\bm{\tilde{a}}^{(1)}$.\\
Solve \eqref{eq:OptProblemMinMax} via \eqref{eq:PowerCond} and \eqref{eq:PowerEq} to attain $\mathcal{P}_{l}, \, \forall l$. \\
\For {$i = 1:L$}{
Find the delay obtained from the Shannon utilization subset of the problem, $\Delta_{i}$. \\
Find semantic delays $\tilde{\Delta}_{l}, \hspace{0.1cm} \forall l$ that achieve better delay than $\Delta_{i}$. \\
Let the subcarriers that satisfy this make a vector $\mathbf{v} = [ m_{1} \hspace{0.05cm} m_{2} \hspace{0.05cm} \cdots \hspace{0.05cm} m_{|M|}]$, where $|M|$ is the number of elements in $\mathbf{v}$ and order is taken with regard to the corresponding delay $\tilde{\Delta}_{m}$. \\
\For{$m = 1:|M|$}{Let $\mathcal{\tilde{P}}_{m}$ be the required power for the $m$-th subcarrier to utilize semantic communication. \\
\If{$\tilde{\Delta}_{m} < \Delta_{i}$ \textup{and}  $\mathcal{P}_{m} > \mathcal{\tilde{P}}_{m}$}
{Set $a_{m} = 0$ and $\tilde{a}_{m} = 1$. \\
\textbf{break}}}
Solve \eqref{eq:OptProblemMinMax} to attain $\mathcal{P}_{l}, \, \forall l$.
} 
Keep last iteration $\bm{a}^{(i)}, \bm{\tilde{a}}^{(i)}$ and $\mathcal{P}_{l}^{(i)}$.
\label{alg:alternateOptMinMax}
\end{algorithm}
Note that the condition $\tilde{\Delta}_{m} < \Delta_{i}$ and  $\mathcal{P}_{m} > \mathcal{\tilde{P}}_{m}$ for each iteration ensures that if a subcarrier prefers to utilize semantic communication due to its lower delay, the remaining power for the subcarriers that utilize Shannon communication will be greater than the previous iteration. As such, the previous power allocation solution is achievable for the remaining subcarriers utilizing Shannon communications and the new power allocation will necessarily achieve better overall delay due to the larger available power. Therefore, the delay of the Shannon utilizing subcarriers will gradually reduce until no further semantic utilization is possible.

The complexity of the proposed algorithm depends on two factors, the first being the number of subcarriers, $L$, used by the system. In the worst case, when all subcarriers must be switched from Shannon to semantic transmission, the algorithm will take at most $L$ iterations to complete. To check the two conditions of the inner for-loop, in the worst case $L$ iterations with two evaluations each are performed, adding $O(2L)$ complexity. The second factor is related to the complexity of solving \eqref{eq:PowerEq}. The complexity of this problem does not depend on the number of subcarriers, but on the number of iterations it takes to converge to a solution below an acceptable threshold. To give a general idea of the complexity and how it affects the proposed algorithm, we denote the complexity of solving \eqref{eq:PowerEq} as $O(X)$ and the complexity of solving problem \eqref{eq:OptProblemMinMax} as $O(|\mathbf{(P4)}|)$. Then, the proposed algorithm must perform $\left(\sfrac{\left(L^2+L\right)}{2}\right)O(X)$ computations because the number of equations required for \eqref{eq:OptProblemMinMax} is reduced by one at each iteration, i.e., $L$ must be solved at the first iteration, $L-1$ at the second iteration, and so on.

Thus, the combined complexity of the proposed algorithm will be equal to
\begin{equation}\label{eq:complexity1}
    O\left(L\log_2L + 2L^2 + \frac{L^2+L}{2}O(X) \right),
\end{equation}
where $O(L\log_2L)$ complexity is required to sort the semantic delays that have been computed before the iterative algorithm.

\section{Optimal Sentence-to-Subcarrier Association}\label{sec:ParallelSem}
Next, we study the existence of an optimal association between sentences' size and subcarrier selection which further minimizes the transmission delay metrics discussed in Section \ref{sec:Opt}.
As stated in Section \ref{sec:SysMod}, due to the strict similarity level of each sentence, all the symbols that compose a sentence have to be transmitted from one subcarrier, while each subcarrier transmits a partition of sentences. Since all sentences $S_{j}$  have in general different character lengths, $8u_{j}$, that implies that the partitions of sentences of all $L$ subcarriers have different data sizes too. Intuitively, to minimize delay, partitions with greater data size should be transmitted by the subcarriers with the best SNR.   

We define serial sentence transmission (SST) using the following channel-sentence correspondence rule
\begin{equation}\label{eq:SSTrule}
    |h_{l}| \gets S_{j}, \,\, j \equiv l\!\!\! \pmod{L},
\end{equation}
where $|h_{l}| \gets S_{j}$ denotes the assignment of sentence $S_{j}$ to the subcarrier with channel $|h_{l}|$.
For instance, for $L=16$ subcarriers and SST, the sentences $S_{1}, S_{23}$ and $S_{75}$ correspond to the subcarriers with channels $|h_{1}|, |h_{7}|$ and $|h_{11}|$, respectively. We notice, though, that SST does not take into account channel magnitudes and sentences' size. Let 
\begin{equation}\label{eq:charlengthOrder}
   u'_{1} \leq \cdots \leq u'_{j} \leq \cdots \leq u'_{NL}
\end{equation}
be the character length of the sentences in ascending order, and without loss of generality, we assume that the ascending order of the channels is given as
\begin{equation}\label{eq:channelOrder}
    |h_{1}| \leq |h_{2}| \leq \cdots \leq |h_{L}|.
\end{equation} 
Furthermore, we define ordered sentence transmission (OST) by the following channel-sentence correspondence rule
\begin{equation}\label{eq:PSTrule}
    |h_{l}| \gets S_{j}, \left\{\forall j | u'_{N(l-1)+1} \leq u'_{j} \leq u'_{Nl} \right\}.
\end{equation}
We note that since $N$ sentences are transmitted to each subcarrier, index $j$ in \eqref{eq:PSTrule} takes $N$ values. In contrast to the SST rule given by \eqref{eq:SSTrule}, the OST mapping in \eqref{eq:PSTrule} takes into account both channel ordering and the amount of data bits of each sentence. Thus, larger sentences are mapped to channels with greater capacities, which can minimize the transmission delay of each subcarrier
Next, we present and prove the following lemma concerning the sum of delays problem.

\begin{lemma}\label{lemma:powerOptimal}
    For \eqref{eq:OptProblemConvex} under the sum-total power equality constraint $\sum_{l=1}^{L} \mathcal{P}_{l} = \mathcal{P}_{\mathrm{tot}}$, regardless of the arrangement of the sentences in the subcarriers, the optimal power distribution,  $\mathcal{P}^{*}$, is obtained so that the capacity of each channel has the same order with the order of the channels.  
\end{lemma}
\begin{IEEEproof}
    The proof is provided in Appendix \ref{app:lemma1}.
\end{IEEEproof}
Algorithm \ref{alg:lemmaOrder} explains the steps described in the proof of Lemma \ref{lemma:powerOptimal}. The key idea is that we divide $Q_i=\mathcal{P}^{*}_{i}|h_i|^2$ by the corresponding channel $h$.

\begin{algorithm}
\caption{Capacity order (Lemma \ref{lemma:powerOptimal})}
For the first repetition, $\mathcal{I}^{(1)}$ are the indices as they appear in \eqref{eq:QCon}. Initial capacity order, $Q_{i_{d}^{(1)}}$, for the assumed optimal power distribution $\mathcal{P}^{*}$. \\
\For {$m = 1:L-1$}{
Find the first out of ascending order index in $\mathcal{I}^{(m)}$. Let this be $i$ and let $j$ be the index that is currently at its position order. Thus, it should hold 
$$i_{d}^{(m+1)} \gets j_{d}^{(m)} = i \text{ and } j_{d}^{(m+1)} \gets i_{d}^{(m)} = j.$$
\\
For the $(m+1)$-th power distribution we change only 
$$(\mathcal{P}^{*}_{i})^{(m+1)} = \frac{Q_{j^{(m)}_{d}}}{|h_{i}|^2} \text{ and } (\mathcal{P}^{*}_{j})^{(m+1)} = \frac{Q_{i^{(m)}_{d}}}{|h_{j}|^2}.$$ \\
The new order of indices $\mathcal{I}^{(m+1)}$ has only two differences from $\mathcal{I}^{(m)}$, but now the largest $m+1$ indices are ordered. 
} 
The last indices set $\mathcal{I}^{(L-1)}$ has all indices in ascending order. $(\mathcal{P}^{*})^{(L-1)}$ is such that
$$\sum_{l=1}^{L} (\mathcal{P}^{*}_{l})^{(L-1)} < \mathcal{P}^{*}.$$
\label{alg:lemmaOrder}
\end{algorithm}

In Lemma \ref{lemma:powerOptimal}, we derived the order of the capacities for the optimal power distribution. In the following lemma, we use this result to obtain the best arrangement of sentences.

\begin{lemma}\label{lemma:Rearrange}
Assuming pure wireless transmission in \eqref{eq:OptProblemConvex} let 
\begin{equation}\label{eq:lengthSerial}
   \mathbf{v}_{1} = \left[ u_{1,1},\cdots, u_{1,L},\cdots, u_{n,l}, \cdots, u_{N,1},\cdots, u_{N,L} \right]
\end{equation}
be the $1 \times NL$ vector consisting of the character lengths of the corresponding sentences as they appear in their serial order and 
\begin{equation}\label{eq:lengthOrder}
    \mathbf{v}_{2} = \left[ u'_{1,1},\cdots, u'_{1,L},\cdots, u'_{n,l}, \cdots, u'_{N,1},\cdots, u'_{N,L} \right]
\end{equation}
the $1 \times NL$ vector consisting of the elements of $\mathbf{v}_{1}$ in ascending order. Then, the ordered sentence transmission is the optimal correspondence between channels and character lengths independently from the serial order of the sentences.
\end{lemma}
\begin{IEEEproof}
The proof is provided in Appendix \ref{app:lemma2}.
\end{IEEEproof}

Utilizing Lemma \ref{lemma:powerOptimal} and Lemma \ref{lemma:Rearrange}, we can state the following theorem.

\begin{theorem}\label{the:OSTOptimal}
    OST is the optimal association of sentences to subcarriers in terms of minimal overall transmission delay time when Shannon wireless transmission is used. 
\end{theorem}

\begin{IEEEproof}
     For the sake of contradiction, we assume that a different arrangement achieves better minimum transmission delay time than OST. From Lemma \ref{lemma:powerOptimal}, the optimal power distribution as obtained from solving \eqref{eq:OptProblemConvex} is such that the capacities of the subcarriers have the same order with their corresponding channels. Then, from Lemma \ref{lemma:Rearrange}, we can show that for this optimal power distribution, OST could achieve reduced transmission delay time if we had arranged differently the sentences in the subcarriers, which contradicts the hypothesis that a different arrangement achieves minimum transmission delay time. Hence, OST is the optimal association policy. 
\end{IEEEproof}

In similar fashion, we present and prove the following theorem concerning the minimization of the maximum delay problem.

\begin{theorem}\label{the:OSTOptimalMinMax}
    OST is the optimal association of sentences to subcarriers in terms of minimal maximum transmission delay time when Shannon wireless transmission is used. 
\end{theorem}

\begin{IEEEproof}
    For the sake of contradiction, let the best possible delay be achieved under some subcarrier-to-subcarrier association different to OST. We assume that for the $m$-th and $l$-th subcarriers $|h_{m}|^2 > |h_{l}|^2$ and $U_{m} < U_{l}$ hold.
    Let the delay achieved by this assignment be $\Delta = D_{m} = D_{l}$ and $\mathcal{P}_{m}, \mathcal{P}_{l}$ be the power resources allocated for their respective subcarriers. Since $U_{m} < U_{l}$ and $D_{m} = D_{l}$ hold, it must be 
    \begin{equation}\label{eq:capOrderMinMax}
        C_{m} < C_{l} \Leftrightarrow \mathcal{P}_{m}|h_{m}|^2 <  \mathcal{P}_{l}|h_{l}|^2.
    \end{equation}
    Then, following Lemma \ref{lemma:powerOptimal}, it is easy to prove that the same delay is possible for OST if the following power resource allocation was performed
    \begin{equation}\label{eq:sysPowerMinMax}
        \mathcal{P}'_{m} = \mathcal{P}_{l}\frac{|h_{l}|^2}{|h_{m}|^2} \text{ and }
        \mathcal{P}'_{l} = \mathcal{P}_{m}\frac{|h_{m}|^2}{|h_{l}|^2}.
    \end{equation}
    Then, for the overall power allocation of the two subcarriers found by \eqref{eq:sysPowerMinMax} it is straightforward that $\mathcal{P}_{m} + \mathcal{P}_{l} > \mathcal{P}'_{m} + \mathcal{P}'_{l}$ holds by the original hypothesis of the channel order and \eqref{eq:capOrderMinMax} in the same way presented in Lemma \ref{lemma:powerOptimal}.
    Therefore, implementing OST can achieve the same delay $\Delta$ as SST allocating less power resources than the latter, meaning that it can achieve better delay than any other sentence-to-subcarrier association. 
\end{IEEEproof}

It is important to note that since we already have an improvement gap of OST over SST and the proposed hybrid system described by \eqref{eq:OptProblemConvex} outperforms the Shannon wireless transmission, we can expect both hybrid policies OST and SST to have better performance than Shannon wireless transmission for both problems under discussion. 

\section{Simulations and Discussion}\label{sec:Results}

\par In this section, simulation results are presented to evaluate the performance of the proposed hybrid semantic-Shannon multi-carrier system. For the channel conditions, we assume Rayleigh fading, thus the channel coefficients follow the circular symmetric complex Gaussian, i.e., $h \sim \mathcal{CN}(0,l_{p})$, where $l_{p} = \left( \frac{\lambda_{c}}{4\pi R} \right)^\nu$ denotes the path loss factor. By $\nu$, $R$, and $\lambda_{c}$ we denote the path loss exponent, the distance between the transmitter and the receiver, and the wavelength of the corresponding central frequency $f_{c}$. 
The bandwidth of each subcarrier is given as $W=\frac{W_{\mathrm{tot}}}{L}$. DeepSC has been pre-trained\footnote{The used text data can be found in \url{http://www.statmt.org/europarl/v7/}.} to get the achievable similarity curve in terms of SNR and $k$ as shown in Fig. \ref{fig:SimilarityTotal}.

All parameters are given in Table \ref{table:Table1}, while all results were averaged over 500 channel realizations and 10 QoS realizations through Monte Carlo simulations. Simulation results address both cases of interest that were discussed in Section \ref{sec:Opt}. From hereon, we refer to problem \eqref{eq:OptProblem} as the Sum problem and \eqref{eq:OptProblemMinMax} as the MinMax problem. To evaluate the performance of the proposed hybrid scheme, we investigate the performance improvement in terms of transmission delay over the Shannon only multi-carrier wireless scheme and the semantic utilization, i.e., the percentage of subcarriers that utilize the DeepSC model. All results are shown with respect to the total average received SNR, which is given as $\mathrm{SNR} = \sfrac{\left(\mathcal{P}_{\mathrm{tot}} \mathbb{E}[|h|^2]\right)}{(N_oW)}$, with $\mathbb{E}[\cdot]$ denoting expectation. The simulation results are derived with no source coding and are performed for three data rates, i.e., the Shannon capacity limit and two uncoded M-QAM schemes with BER-achieving thresholds $10^{-3}$ and $10^{-5}$.

\begin{table}[t] 
    \centering
    \caption{Simulation Parameters}
    \label{table:Table1}
    \begin{tabular}{|c|c|}
    \hline
    \textbf{Parameter} & \textbf{Value (Unit)} \\
    \hline \hline
    Noise power spectral density, $N_0$ & $-174$ dBm/Hz \\
    \hline
    Total bandwidth, $W_{\mathrm{tot}}$ & $20$ MHz \\
    \hline
    Frequency, $f_c$ & $2.4$ GHz \\
    \hline
    Distance, $R$ & $100$ m \\
    \hline
    Path loss exponent, $\nu$ & $2$ \\
    \hline
    Semantic symbols, $k$ & $16$ symbols/word \\
    \hline
    Similarity threshold, $M_{n,l}^{\mathrm{th}}$ & $[0.6,1]$ \\
    \hline
    Similarity Upper Bound, $M_{\mathrm{sat}}$ & $0.98$ \\
    \hline
    Sentence Length, $L_j$ & $4 - 32$ \\
    \hline
    Number of subcarriers, $L$ & $64$ \\
    \hline  
    Number of sentences, $P$ & $7296$ \\
    \hline
    \end{tabular} 
\end{table}

\begin{figure}[t]
    \centering
    \begin{tikzpicture}
        \begin{axis}[
            width=.9\linewidth,
            xlabel = {Total average received SNR (dB)},
            ylabel = {Semantic utilization $(\%)$},
            ymin = 0,
            ymax = 72,
            xmin = 10,
            xmax = 35,
            ytick = {0,10,20,30,40,50,60,70},
            yticklabels = {0,10,20,30,40,50,60,70},
            grid = major,
            legend columns=2, 
		legend entries ={{SST},{OST},{$L=16$},{$L=64$},{$L=128$}},
            legend cell align = {left},
            legend style={font=\scriptsize},
            legend style={at={(1,1)},anchor=north east},
            ]
            \addplot[
            blue,
            mark = none,
            line width = 1pt,
            style = solid,
            ]
            table {Data/Util_k16_SST/Util_k16_SST1.dat};
            \addplot[
            red,
            mark = none,
            line width = 1pt,
            style = solid,
            ]
            table {Data/Util_k16_OST/Util_k16_OST1.dat};
            \addplot[
            black,
            mark = square,
            mark repeat = 2,
            mark size = 2,
            only marks,
            ]
            table {Data/Util_k16_SST/Util_k16_SST1.dat};
            \addplot[
            black,
            mark = o,
            mark repeat = 2,
            mark size = 2,
            only marks,
            ]
            table {Data/Util_k16_SST/Util_k16_SST3.dat};
            \addplot[
            black,
            mark = triangle,
            mark repeat = 2,
            mark size = 2,
            only marks,
            ]
            table {Data/Util_k16_SST/Util_k16_SST4.dat};
            \addplot[
            blue,
            mark = square,
            mark repeat = 2,
            mark size = 2,
            only marks,
            ]
            table {Data/Util_k16_SST/Util_k16_SST1.dat};
            \addplot[
            blue,
            mark = o,
            mark repeat = 2,
            mark size = 2,
            only marks,
            ]
            table {Data/Util_k16_SST/Util_k16_SST3.dat};
            \addplot[
            blue,
            mark = triangle,
            mark repeat = 2,
            mark size = 2,
            only marks,
            ]
            table {Data/Util_k16_SST/Util_k16_SST4.dat};
            \addplot[
            blue,
            mark = none,
            line width = 1pt,
            style = solid,
            ]
            table {Data/Util_k16_SST/Util_k16_SST3.dat};
            \addplot[
            blue,
            mark = none,
            line width = 1pt,
            style = solid,
            ]
            table {Data/Util_k16_SST/Util_k16_SST4.dat};
            \addplot[
            red,
            mark = square,
            mark repeat = 2,
            mark size = 2,
            only marks,
            ]
            table {Data/Util_k16_OST/Util_k16_OST1.dat};
            \addplot[
            red,
            mark = o,
            mark repeat = 2,
            mark size = 2,
            line width = 1pt,
            style = solid,
            ]
            table {Data/Util_k16_OST/Util_k16_OST3.dat};
            \addplot[
            red,
            mark = triangle,
            mark repeat = 2,
            mark size = 2,
            line width = 1pt,
            style = solid,
            ]
            table {Data/Util_k16_OST/Util_k16_OST4.dat};
        \end{axis}
    \end{tikzpicture}
    \caption{Semantic utilization of the Sum problem for $k=16$ DNN outputs.}
    \label{fig:Utilk16}
\end{figure}
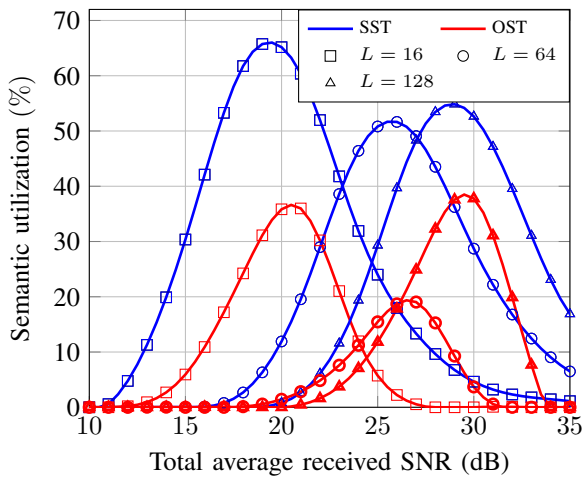

\begin{figure}[t]
    \centering
    \begin{tikzpicture}
        \begin{axis}[
            width=.9\linewidth,
            xlabel = {Total average received SNR (dB)},
            ylabel = {Semantic utilization $(\%)$},
            ymin = 0,
            ymax = 50,
            xmin = 15,
            xmax = 35,
            ytick = {0,10,20,30,40,50},
            yticklabels = {0,10,20,30,40,50},
            grid = major,
            legend columns=2, 
		legend entries ={{SST},{OST},{$L=16$},{$L=64$},{$L=128$}},
            legend cell align = {left},
            legend style={font=\scriptsize},
            legend style={at={(1,1)},anchor=north east},
            ]
            \addplot[
            blue,
            mark = none,
            line width = 1pt,
            style = solid,
            ]
            table {Data/MinMax_k16_SST/MinMax_Util_L16_SST.dat};
            \addplot[
            red,
            mark = none,
            line width = 1pt,
            style = solid,
            ]
            table {Data/MinMax_k16_OST/MinMax_Util_L16_OST.dat};
            \addplot[
            black,
            mark = square,
            mark repeat = 2,
            mark size = 2,
            only marks,
            ]
            table {Data/MinMax_k16_SST/MinMax_Util_L16_SST.dat};
            \addplot[
            black,
            mark = o,
            mark repeat = 2,
            mark size = 2,
            only marks,
            ]
            table {Data/MinMax_Util/Util_MinMax.dat};
            \addplot[
            black,
            mark = triangle,
            mark repeat = 2,
            mark size = 2,
            only marks,
            ]
            table {Data/MinMax_k16_SST/MinMax_Util_L128_SST.dat};
            \addplot[
            blue,
            mark = square,
            mark repeat = 2,
            mark size = 2,
            only marks,
            ]
            table {Data/MinMax_k16_SST/MinMax_Util_L16_SST.dat};
            \addplot[
            blue,
            mark = o,
            mark repeat = 2,
            mark size = 2,
            only marks,
            ]
            table {Data/MinMax_Util/Util_MinMax.dat};
            \addplot[
            blue,
            mark = triangle,
            mark repeat = 2,
            mark size = 2,
            only marks,
            ]
            table {Data/MinMax_k16_SST/MinMax_Util_L128_SST.dat};
            \addplot[
            blue,
            mark = none,
            line width = 1pt,
            style = solid,
            ]
            table {Data/MinMax_Util/Util_MinMax.dat};
            \addplot[
            blue,
            mark = none,
            line width = 1pt,
            style = solid,
            ]
            table {Data/MinMax_k16_SST/MinMax_Util_L128_SST.dat};
            \addplot[
            red,
            mark = square,
            mark repeat = 2,
            mark size = 2,
            only marks,
            ]
            table {Data/MinMax_k16_OST/MinMax_Util_L16_OST.dat};
            \addplot[
            red,
            mark = o,
            mark repeat = 2,
            mark size = 2,
            line width = 1pt,
            style = solid,
            ]
            table {Data/MinMax_k16_OST/MinMax_Util_L64_OST.dat};
            \addplot[
            red,
            mark = triangle,
            mark repeat = 2,
            mark size = 2,
            line width = 1pt,
            style = solid,
            ]
            table {Data/MinMax_k16_OST/MinMax_Util_L128_OST.dat};
        \end{axis}
    \end{tikzpicture}
    \caption{Semantic utilization of the MinMax problem for $k=16$ DNN outputs.}
    \label{fig:utilMinMaxk16}
\end{figure}
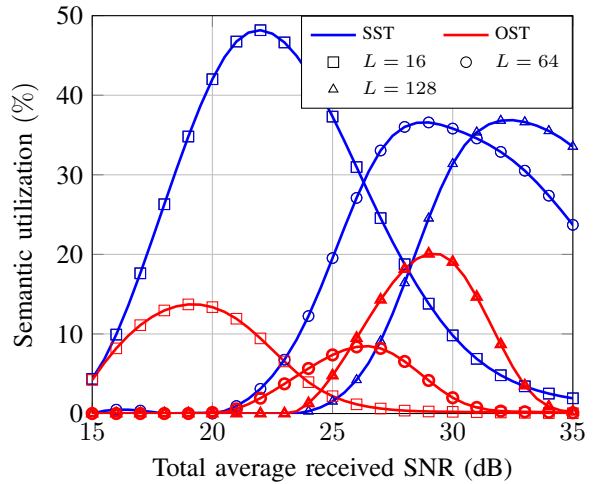

\begin{figure}[t]
    \centering
    \begin{tikzpicture}
        \begin{axis}[
            width=.9\linewidth,
            xlabel = {Total average received SNR (dB)},
            ylabel = {Semantic utilization $(\%)$},
            ymin = 0,
            ymax = 80,
            xmin = 15,
            xmax = 35,
            ytick = {0,10,20,30,40,50,60,70,80},
            yticklabels = {0,10,20,30,40,50,60,70,80},
            grid = major,
		legend entries ={{Sum},{MinMax},{Shannon},{$\mathrm{BER}=10^{-3}$},{$\mathrm{BER}=10^{-5}$}},
            legend cell align = {left},
            legend style={font=\footnotesize},
            legend style={at={(0,1)},anchor=north west},
            ]
            \addplot[
            blue,
            mark = none,
            line width = 1pt,
            style = solid,
            ]
            table {Data/Util_k16_SST/Util_k16_SST3.dat};
            \addplot[
            red,
            mark = none,
            line width = 1pt,
            style = solid,
            ]
            table {Data/MinMax_Util/Util_MinMax.dat};
            \addplot[
            black,
            mark = square,
            mark repeat = 2,
            mark size = 2,
            only marks,
            ]
            table {Data/Util_k16_SST/Util_k16_SST3.dat};
            \addplot[
            black,
            mark = o,
            mark repeat = 2,
            mark size = 2,
            only marks,
            ]
            table {Data/Util_BER/Util_BER1.dat};
            \addplot[
            black,
            mark = triangle,
            mark repeat = 2,
            mark size = 2,
            only marks,
            ]
            table {Data/Util_BER/Util_BER3.dat};
            \addplot[
            blue,
            mark = square,
            mark repeat = 2,
            mark size = 2,
            only marks,
            ]
            table {Data/Util_k16_SST/Util_k16_SST3.dat};
            \addplot[
            blue,
            mark = o,
            mark repeat = 2,
            mark size = 2,
            only marks,
            ]
            table {Data/Util_BER/Util_BER1.dat};
            \addplot[
            blue,
            mark = triangle,
            mark repeat = 2,
            mark size = 2,
            only marks,
            ]
            table {Data/Util_BER/Util_BER3.dat};
            \addplot[
            blue,
            mark = none,
            line width = 1pt,
            style = solid,
            ]
            table {Data/Util_BER/Util_BER1.dat};
            \addplot[
            blue,
            mark = none,
            line width = 1pt,
            style = solid,
            ]
            table {Data/Util_BER/Util_BER3.dat};
            \addplot[
            red,
            mark = square,
            mark repeat = 2,
            mark size = 2,
            only marks,
            ]
            table {Data/MinMax_Util/Util_MinMax.dat};
            \addplot[
            red,
            mark = o,
            mark repeat = 2,
            mark size = 2,
            line width = 1pt,
            style = solid,
            ]
            table {Data/MinMax_Util/MinMax_Util_BER1.dat};
            \addplot[
            red,
            mark = triangle,
            mark repeat = 2,
            mark size = 2,
            line width = 1pt,
            style = solid,
            ]
            table {Data/MinMax_Util/MinMax_Util_BER2.dat};
        \end{axis}
    \end{tikzpicture}
    \caption{SST semantic utilization for $k=16$ DNN outputs.}
    \label{fig:utilBER}
\end{figure}
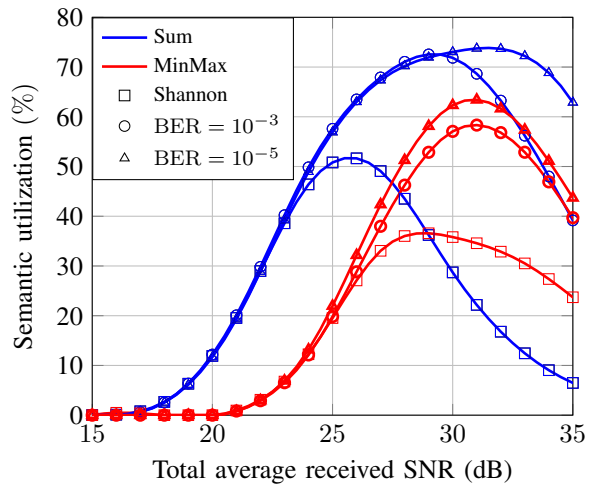

In Fig. \ref{fig:Utilk16}, the semantic utilization for both SST and OST is plotted for the Sum problem. We observe that the semantic utilization reaches its peak for medium SNR values, while reaching zero in the low and high SNR regime, implying that only Shannon communications are used. This is attributed, first, to the fact that in the low SNR regime, DeepSC cannot achieve the required similarity, since each subcarrier has not the desired transmit SNR, while in the high SNR regime, the transmission rate of DeepSC is smaller than the Shannon digital communication scheme. We note that the transmission rate of the DeepSC model is fixed and equal to its available bandwidth, while the transmission rate of digital communications increases as the available transmit power increases. As such, semantic transmission based on DeepSC is preferable only in the medium SNR regime. Also, the SNR regime, where the semantic utilization is maximum, is shifted to the right as the number of subcarriers increases, since more transmit power is needed to guarantee the semantic QoS at all subcarriers. It is notable that the semantic utilization can take values between 50-60\%, implying that half the subcarriers utilize DeepSC. With regards to OST it is observed that semantic utilization has considerably smaller values compared to SST,  around 20-50\%. According to \eqref{eq:maxSim}, the semantic QoS must hold for all sentences associated with a subcarrier inside the same coherence time interval, thus the maximum of all the semantic QoS has to be guaranteed. The OST is based on reordering the sentences and then associating them with subcarriers. This reordering affects the distribution of the maximum QoS which is associated with each subcarrier. Therefore, in OST, it is likely that a subcarrier will be associated to at least one sentence which has a similarity threshold greater than $M_\mathrm{sat}$, and as a consequence, that subcarrier is forced to use Shannon communication. 

In Fig. \ref{fig:utilMinMaxk16} the semantic utilization of both SST and OST is plotted for the MinMax problem. In similar fashion to Fig. \ref{fig:Utilk16}, it is observed that for increasing number of subcarriers, $L$, semantic utilization is decreased and occurs for greater values of total average received SNR due to the joint power resources of the system. Furthermore, as in the Sum problem, the semantic utilization of OST is smaller compared to that of SST as previously explained in Fig. \ref{fig:Utilk16}. Moreover, it is important to highlight that the utilization levels achieved in the MinMax problem are smaller than that of the Sum problem, which is an inherent characteristic of the MinMax problem, because, as showcased by Algorithm \ref{alg:alternateOptMinMax}, subcarriers can utilize semantic communications only as long as the transmission delay of the subcarriers utilizing Shannon communications is greater than the transmission delay achieved by semantic utilization. This restriction is not necessary for the Sum problem, thus in the latter semantic communications can be more widely utilized. However, it is worth noting that in the MinMax problem semantic communications can be utilized even in higher SNR values, because subcarriers that can achieve considerably small transmission delay will favor semantic utilization over Shannon until the latter is preferable for each subcarrier. 

In Fig. \ref{fig:utilBER}, the semantic utilization for SST is plotted under different BER thresholds. It is noted that when BER is taken into account, the semantic communications utilization increases, due to the fact that BER limits the maximum achievable data rate of digital communication as shown in \eqref{eq:Shannon}. Nevertheless, in the low and high SNR regimes, the semantic utilization again drops towards zero. Therefore, semantic communications can be used to reduce the overall transmission delay, especially in the medium SNR regime. We note that this behavior is similar to the one under the assumption of capacity achieving data rate transmission, since increasing the SNR eventually allows Shannon communications to outperform the delay achieved by semantic communications. 

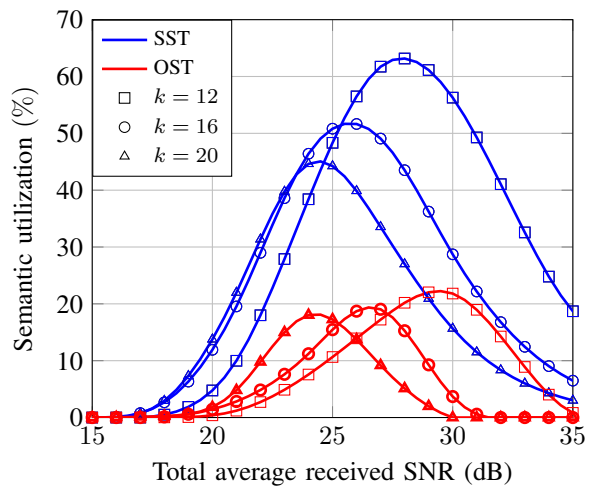
\begin{figure}[ht]
    \centering
    \begin{tikzpicture}
        \begin{axis}[
            width=.9\linewidth,
            xlabel = {Total average received SNR (dB)},
            ylabel = {Semantic utilization $(\%)$},
            ymin = 0,
            ymax = 70,
            xmin = 15,
            xmax = 35,
            ytick = {0,10,20,30,40,50,60,70},
            yticklabels = {0,10,20,30,40,50,60,70},
            grid = major,
		legend entries ={{SST},{OST},{$k=12$},{$k=16$},{$k=20$}},
            legend cell align = {left},
            legend style={font=\footnotesize},
            legend style={at={(0,1)},anchor=north west},
            ]
            \addplot[
            blue,
            mark = none,
            line width = 1pt,
            style = solid,
            ]
            table {Data/Util_L64_SST/Util_L64_SST2.dat};
            \addplot[
            red,
            mark = none,
            line width = 1pt,
            style = solid,
            ]
            table {Data/Util_L64_OST/Util_L64_OST2.dat};
            \addplot[
            black,
            mark = square,
            mark repeat = 2,
            mark size = 2,
            only marks,
            ]
            table {Data/Util_L64_SST/Util_L64_SST2.dat};
            \addplot[
            black,
            mark = o,
            mark repeat = 2,
            mark size = 2,
            only marks,
            ]
            table {Data/Util_L64_SST/Util_L64_SST4.dat};
            \addplot[
            black,
            mark = triangle,
            mark repeat = 2,
            mark size = 2,
            only marks,
            ]
            table {Data/Util_L64_SST/Util_L64_SST6.dat};
            \addplot[
            blue,
            mark = square,
            mark repeat = 2,
            mark size = 2,
            only marks,
            ]
            table {Data/Util_L64_SST/Util_L64_SST2.dat};
            \addplot[
            blue,
            mark = o,
            mark repeat = 2,
            mark size = 2,
            only marks,
            ]
            table {Data/Util_L64_SST/Util_L64_SST4.dat};
            \addplot[
            blue,
            mark = triangle,
            mark repeat = 2,
            mark size = 2,
            only marks,
            ]
            table {Data/Util_L64_SST/Util_L64_SST6.dat};
            \addplot[
            blue,
            mark = none,
            line width = 1pt,
            style = solid,
            ]
            table {Data/Util_L64_SST/Util_L64_SST4.dat};
            \addplot[
            blue,
            mark = none,
            line width = 1pt,
            style = solid,
            ]
            table {Data/Util_L64_SST/Util_L64_SST6.dat};
            \addplot[
            red,
            mark = square,
            mark repeat = 2,
            mark size = 2,
            only marks,
            ]
            table {Data/Util_L64_OST/Util_L64_OST2.dat};
            \addplot[
            red,
            mark = o,
            mark repeat = 2,
            mark size = 2,
            line width = 1pt,
            style = solid,
            ]
            table {Data/Util_L64_OST/Util_L64_OST4.dat};
            \addplot[
            red,
            mark = triangle,
            mark repeat = 2,
            mark size = 2,
            line width = 1pt,
            style = solid,
            ]
            table {Data/Util_L64_OST/Util_L64_OST6.dat};
        \end{axis}
    \end{tikzpicture}
    \caption{Semantic utilization of the Sum problem for $L=64$ DNN outputs.}
    \label{fig:UtilL64}
\end{figure}

In Fig. \ref{fig:UtilL64}, the semantic utilization of both SST and OST protocols for the Sum problem and various number of semantic symbols per word values are illustrated. As $k$ decreases, the achievable similarity also decreases, thus $k<12$ was not investigated, since DeepSC is rarely selected in this case. First, it is observed that for greater values of $k$ the peak of the semantic utilization arises in smaller SNR values, which is due to greater values of $k$ providing increased similarity for smaller SNR values. However, the semantic utilization decreases as $k$ increases, since more semantic symbols need to be transmitted. By taking into account that DeepSC has smaller transmission rate than Shannon communication, this increase in semantic symbols per word results in increased transmission delay when DeepSC is employed, thus Shannon communication is preferred. It is important to note that for both SST and OST the same intuitions hold regardless of the number of DDN outputs, $k$, although, the semantic utilization values for OST are smaller than SST, which was explained in Fig. \ref{fig:Utilk16}.

\begin{figure}[ht]
    \centering
    \begin{tikzpicture}
        \begin{axis}[
            width=.9\linewidth,
            xlabel = {Total average received SNR (dB)},
            ylabel = {Transmission delay improvement $(\%)$},
            ymin = 0,
            ymax = 80,
            xmin = 10,
            xmax = 35,
            ytick = {0,10,20,30,40,50,60,70,80},
            yticklabels = {0,10,20,30,40,50,60,70,80},
            grid = major,
		legend entries ={{Sum},{MinMax},{Hybrid SST},{Shannon OST},{Hybrid OST}},
            legend cell align = {left},
            legend style={font=\scriptsize},
            legend style={at={(1,1)},anchor=north east},
            ]
            \addplot[
            blue,
            mark = none,
            line width = 1pt,
            style = solid,
            ]
            table {Data/Imp_L32/Imp_L32_1.dat};
            \addplot[
            red,
            mark = none,
            line width = 1pt,
            style = solid,
            ]
            table {Data/ImpL32/Imp_OST1.dat};
            \addplot[
            black,
            mark = square,
            mark repeat = 2,
            mark size = 2,
            only marks,
            ]
            table {Data/Imp_L32/Imp_L32_1.dat};
            \addplot[
            black,
            mark = o,
            mark repeat = 2,
            mark size = 2,
            only marks,
            ]
            table {Data/Imp_L32/Imp_L32_2.dat};
            \addplot[
            black,
            mark = triangle,
            mark repeat = 2,
            mark size = 2,
            only marks,
            ]
            table {Data/Imp_L32/Imp_L32_3.dat};
            \addplot[
            blue,
            mark = square,
            mark repeat = 2,
            mark size = 2,
            only marks,
            ]
            table {Data/Imp_L32/Imp_L32_1.dat};
            \addplot[
            blue,
            mark = o,
            mark repeat = 2,
            mark size = 2,
            only marks,
            ]
            table {Data/Imp_L32/Imp_L32_2.dat};
            \addplot[
            blue,
            mark = triangle,
            mark repeat = 2,
            mark size = 2,
            only marks,
            ]
            table {Data/Imp_L32/Imp_L32_3.dat};
            \addplot[
            blue,
            mark = none,
            line width = 1pt,
            style = solid,
            ]
            table {Data/Imp_L32/Imp_L32_2.dat};
            \addplot[
            blue,
            mark = none,
            line width = 1pt,
            style = solid,
            ]
            table {Data/Imp_L32/Imp_L32_3.dat};
            \addplot[
            red,
            mark = square,
            mark repeat = 2,
            mark size = 2,
            only marks,
            ]
            table {Data/ImpL32/Imp_OST1.dat};
            \addplot[
            red,
            mark = o,
            mark repeat = 2,
            mark size = 2,
            line width = 1pt,
            style = solid,
            ]
            table {Data/ImpL32/Imp_OST2.dat};
            \addplot[
            red,
            mark = triangle,
            mark repeat = 2,
            mark size = 2,
            line width = 1pt,
            style = solid,
            ]
            table {Data/ImpL32/Imp_OST3.dat};
        \end{axis}
    \end{tikzpicture}
    \caption{Transmission delay improvement of the Sum problem for $L=32$ subcarriers and $k=16$.}
    \label{fig:gainL32}
\end{figure}
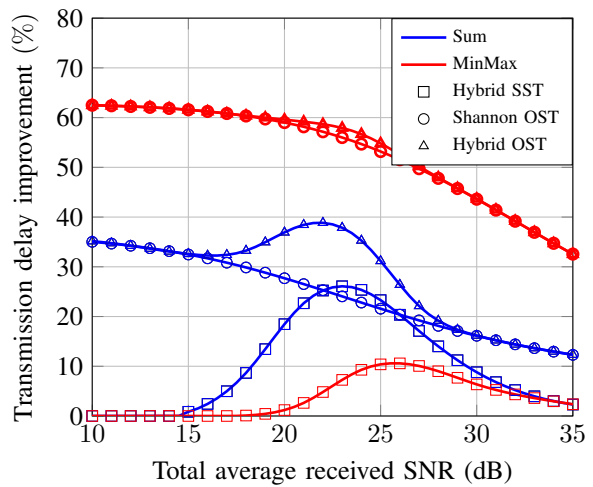

 In Fig. \ref{fig:gainL32} the performance improvement, in terms of transmission delay, of the proposed schemes against the Shannon SST scheme for the Sum problem is illustrated. First, it is observed that the OST protocol is always better than the SST protocol, which numerically validates the analysis of Section \ref{sec:ParallelSem} and Theorems \ref{the:OSTOptimal} and \ref{the:OSTOptimalMinMax}. The improvement diminishes as the SNR increases, since as the transmission power increases the impact of the sentences' size on the delay reduces. However, in the low SNR regime where the decisive factor is the sentences' size, the OST protocols achieve their maximum improvement. This further confirms the importance of the OST protocol, regardless if it is used in conjunction with semantic communications. The effectiveness of the OST is also proven by observing that the Shannon OST scheme almost always outperforms the hybrid SST scheme. Thus, the delay decrease provided by the reordering of the sentences is in general greater than the delay decrease provided by using semantic communications. By integrating the OST protocol with semantic transmission, the hybrid scheme achieves superior performance compared to all considered communication schemes, since it combines the advantages of both OST and semantic communications. It is worth noting that the implementation of the OST protocol for the MinMax problem achieves an improvement around $60\%$ over the SST protocol counterparts, which highlights the significance of the proposed sentence-to-subcarrier association.

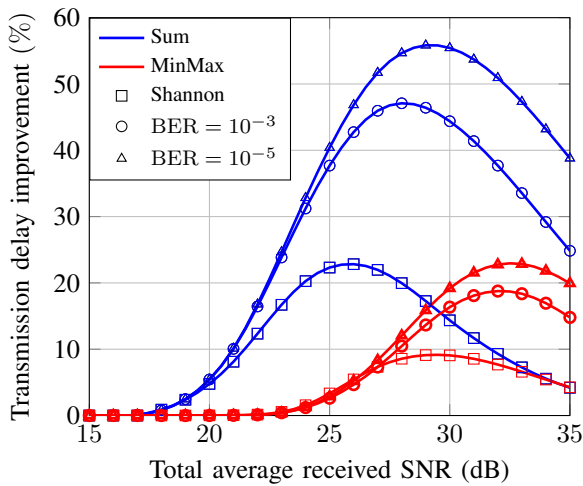
\begin{figure}[ht]
    \centering
    \begin{tikzpicture}
        \begin{axis}[
            width=.9\linewidth,
            xlabel = {Total average received SNR (dB)},
            ylabel = {Transmission delay improvement $(\%)$},
            ymin = 0,
            ymax = 60,
            xmin = 15,
            xmax = 35,
            ytick = {0,10,20,30,40,50,60},
            yticklabels = {0,10,20,30,40,50,60},
            grid = major,
		legend entries ={{Sum},{MinMax},{Shannon},{$\mathrm{BER}=10^{-3}$},{$\mathrm{BER}=10^{-5}$}},
            legend cell align = {left},
            legend style={font=\footnotesize},
            legend style={at={(0,1)},anchor=north west},
            ]
            \addplot[
            blue,
            mark = none,
            line width = 1pt,
            style = solid,
            ]
            table {Data/Imp_L64/Imp_L64_1.dat};
            \addplot[
            red,
            mark = none,
            line width = 1pt,
            style = solid,
            ]
            table {Data/MinMax_Imp/Imp_MinMax.dat};
            \addplot[
            black,
            mark = square,
            mark repeat = 2,
            mark size = 2,
            only marks,
            ]
            table {Data/Imp_L64/Imp_L64_1.dat};
            \addplot[
            black,
            mark = o,
            mark repeat = 2,
            mark size = 2,
            only marks,
            ]
            table {Data/Imp_BER/Imp_BER1.dat};
            \addplot[
            black,
            mark = triangle,
            mark repeat = 2,
            mark size = 2,
            only marks,
            ]
            table {Data/Imp_BER/Imp_BER3.dat};
            \addplot[
            blue,
            mark = square,
            mark repeat = 2,
            mark size = 2,
            only marks,
            ]
            table {Data/Imp_L64/Imp_L64_1.dat};
            \addplot[
            blue,
            mark = o,
            mark repeat = 2,
            mark size = 2,
            only marks,
            ]
            table {Data/Imp_BER/Imp_BER1.dat};
            \addplot[
            blue,
            mark = triangle,
            mark repeat = 2,
            mark size = 2,
            only marks,
            ]
            table {Data/Imp_BER/Imp_BER3.dat};
            \addplot[
            blue,
            mark = none,
            line width = 1pt,
            style = solid,
            ]
            table {Data/Imp_BER/Imp_BER1.dat};
            \addplot[
            blue,
            mark = none,
            line width = 1pt,
            style = solid,
            ]
            table {Data/Imp_BER/Imp_BER3.dat};
            \addplot[
            red,
            mark = square,
            mark repeat = 2,
            mark size = 2,
            only marks,
            ]
            table {Data/MinMax_Imp/Imp_MinMax.dat};
            \addplot[
            red,
            mark = o,
            mark repeat = 2,
            mark size = 2,
            line width = 1pt,
            style = solid,
            ]
            table {Data/MinMax_Imp/MinMax_Imp_BER1.dat};
            \addplot[
            red,
            mark = triangle,
            mark repeat = 2,
            mark size = 2,
            line width = 1pt,
            style = solid,
            ]
            table {Data/MinMax_Imp/MinMax_Imp_BER2.dat};
        \end{axis}
    \end{tikzpicture}
    \caption{SST transmission delay improvement for $L=64$ subcarriers and $k=16$ DNN outputs.}
    \label{fig:gainBER}
\end{figure}

\par In Fig. \ref{fig:gainBER}, the transmission delay improvement of both problems for SST and different BER thresholds is presented. The improvement is associated with the semantic utilization, since the larger the latter is, the larger the improvement will be. However, it is observed that the transmission delay improvement has a similar behavior to that of Shannon's capacity case, because the same data rates will eventually be achieved by the uncoded $M$-QAM schemes, but for greater values of SNR. This ensures that at some point the transmission delay time of each subcarrier under Shannon communications will become smaller that the one achieved by the semantic communications and, thus, as suggested by Fig. \ref{fig:utilBER}, the latter will not provide any improvement. As observed, both problems are characterized by similar behavior for increasing data rates, however the MinMax problem achieves smaller improvement due to the smaller semantic utilization compared to the Sum problem as illustrated in Figs. \ref{fig:utilMinMaxk16} and \ref{fig:utilBER}.

It is worth noting that since the Shannon capacity limit is the maximum achievable rate, any other practical coded scheme will achieve performance between that achieved by Shannon capacity and the lower uncoded BER threshold. This implies that since semantic utilization leads to performance improvement even in the extreme case of Shannon-achieving coding, semantic utilization will also be preferable for all other schemes. The increase in semantic utilization as the BER threshold decreases in Fig. \ref{fig:utilBER} also illustrates this point, thus demonstrating the importance of integrating semantic communication along with conventional Shannon communication.

\section{Conclusions}\label{sec:conclusions}
In this paper, the synergy of Shannon wireless communications with state-of-the-art semantic communication systems for text transmission was examined. The minimization of two transmission delay metrics of the proposed hybrid system was investigated,  subject to strict similarity levels between the original and the reconstructed data. Furthermore, we proved that for the proposed multi-carrier hybrid system, the arrangement of data for transmission significantly affects the performance of the system and the optimal association between sentences and subcarriers was provided. The simulation results illustrate that semantic communications are not always the preferable way of transmission and Shannon communications still achieve better transmission delay, even for non capacity-achieving data rates, for specific SNR regimes. Finally, we concluded that the utilization of semantic communications can decrease the transmission delay time of the system, but further research is needed to unveil the advantages and disadvantages of semantic communications.  
\appendices

\section{Proof of Lemma \ref{lemma:powerOptimal}} \label{app:lemma1}
    Without loss of generality, we assume that \eqref{eq:channelOrder} gives the channels' magnitude in ascending order. For the sake of contradiction, we suppose that the optimal power distribution $\mathcal{P}^{*}$ is such that the channels' magnitude order given in \eqref{eq:channelOrder} and the capacity order corresponding to this power distribution are in different order, for instance, 
    \begin{equation}\label{eq:capacitiesCon}
    C_{1} \leq  \cdots \leq C_{i} \leq \cdots \leq C_{j} \leq \cdots \leq C_{L},
    \end{equation}
    while $h_{i},h_{j}$ are such that $|h_{j}| \leq |h_{i}|$. For convenience, we denote $Q_{i} = \mathcal{P}^{*}_{i}|h_i|^2$. Then, \eqref{eq:capacitiesCon} is equivalent to
    \begin{equation}\label{eq:QCon}
    Q_{1} \leq  \cdots \leq Q_{i} \leq \cdots \leq Q_{j} \leq \cdots \leq Q_{L}.
    \end{equation}
    
    \par We will prove that there exists a different power distribution which can achieve the same capacity values, but with less power consumption. To prove this, we will show that the capacity ordering that leads to minimum power consumption is the same as the ordering of the channels associated with the capacities, i.e., if $|h_1|<|h_2|$, the optimal ordering is $Q_1 < Q_2$. 

    \par Let $(\mathcal{P}^{*})^{(m)}$ be the $m$-th power distribution given by 
    \begin{equation} \label{eq:power1}      (\mathcal{P}^{*}_{i})^{(m)} = \frac{Q_{i^{(m)}_{d}}}{|h_i|^2} ,
    \end{equation}
    where $i^{(m)}_{d}$ symbolizes the destination index of the $i$-th channel in the $m$-th repetition and the last $m$ terms are in ascending order. Since only two capacities are ordered in each repetition, only two values change from one power distribution to the next. Let $i$ be the $m$-th out of order capacity and $j$ be currently in its order instead. Therefore, it is $i^{(m+1)}_{d} = j^{(m)}_{d} = i$ and, since they swap places, it also is $j^{(m+1)}_{d} = i^{(m)}_{d}=j$. Then, we have that  
    \begin{equation}\label{eq:power2}
        (\mathcal{P}^{*}_{i})^{(m)} = \frac{Q_{i^{(m)}_{d}}}{|h_{i}|^2} \text{ and } (\mathcal{P}^{*}_{j})^{(m)} = \frac{Q_{j^{(m)}_{d}}}{|h_{j}|^2}
    \end{equation}
    and we can observe that by choosing 
    \begin{equation}\label{eq:power3}
        (\mathcal{P}^{*}_{i})^{(m+1)} = \frac{Q_{j^{(m)}_{d}}}{|h_{i}|^2} \text{ and } (\mathcal{P}^{*}_{j})^{(m+1)} = \frac{Q_{i^{(m)}_{d}}}{|h_{j}|^2},
    \end{equation}
    we would get $C_{i}^{(m+1)} = C_{j}$, $C_{j}^{(m+1)} = C_{i}$ and \eqref{eq:capacitiesCon} would be satisfied. However, the power distribution $(\mathcal{P}^{*})^{(m+1)}$ achieves better power consumption, because
    \begin{equation}\label{eq:powerFeasible1}
         (\mathcal{P}^{*}_{j})^{(m+1)} + (\mathcal{P}^{*}_{i})^{(m+1)} < (\mathcal{P}^{*}_{j})^{(m)} + (\mathcal{P}^{*}_{i})^{(m)} ,
    \end{equation}
    which is equivalent to 
    \begin{equation}\label{eq:powerFeasible2}\left[ Q_{j^{(m)}_{d}} - Q_{i^{(m)}_{d}} \right] \left( \frac{1}{|h_i|^2} - \frac{1}{|h_j|^2}\right) < 0
    \end{equation}
    and \eqref{eq:powerFeasible2} holds by channel ordering and the order given in \eqref{eq:QCon}. It should be highlighted that due to \eqref{eq:powerFeasible2} the power measures described in \eqref{eq:power3} are feasible, because they lead to lower power consumption without violating the sum-total power constraint of \eqref{eq:OptProblemConvex}. 
    \par The same process can be repeated until all capacities are in the same order with their corresponding channels, which takes at most $L-1$ repetitions. At the last repetition, the current power distribution $(\mathcal{P}^{*})^{(L-1)}$ is such that 
    \begin{equation}\label{eq:CapacityRepOrder}
        \left[C_{1}^{(L-1)}, C_{2}^{(L-1)}, \cdots, C_{L}^{(L-1)} \right] = \left[ C_{1}, C_{2}, \cdots, C_{L} \right],
    \end{equation}
    where the vector on the right-hand side of \eqref{eq:CapacityRepOrder} is the initial order given by \eqref{eq:capacitiesCon} and the left-hand side consists of the same capacities, but in ascending order with respect to their corresponding channels, and is achievable directly with less power consumption than the assumed optimal power distribution. Thus, the initial found power distribution $\mathcal{P}^{*}$ cannot be the optimal power distribution, contradicting the original assumption. 

\section{Proof of Lemma \ref{lemma:Rearrange}} \label{app:lemma2}
Without loss of generality, we assume that the channels' magnitudes in ascending order are given by \eqref{eq:channelOrder} and that $\mathcal{P}^{*}$ is the optimal power distribution which minimizes the transmission delay time of $\mathbf{v}_{1}$.
Then, by Lemma \ref{lemma:powerOptimal}, for the  power distribution $\mathcal{P}^{*}$ it should hold that
\begin{equation}\label{eq:rateOrder}
    \frac{1}{C_{1}} \geq \frac{1}{C_{2}} \geq \cdots \geq \frac{1}{C_{L}},
\end{equation}
due to  the logarithm being an increasing function and the capacities having the same order with their corresponding channels. We define the $1 \times NL$ vector
\begin{equation}\label{eq:rateTuple}
    \mathbf{v}_{3} = \left[\underbrace{\frac{1}{C_{1}},\cdots, \frac{1}{C_{1}}}_{N \text{ times}},\cdots, \underbrace{\frac{1}{C_{l}}, \cdots, \frac{1}{C_{l}}}_{N \text{ times}}, \cdots, \underbrace{\frac{1}{C_{L}},\cdots, \frac{1}{C_{L}}}_{N \text{ times}} \right],
\end{equation}
the elements of which is in descending order. Then, by the rearrangement inequality \cite{cvetkovski2012inequalities}, and since $\mathbf{v}_{2}$ and $\mathbf{v}_{3}$ are oppositely sorted, it will be
\begin{equation}\label{eq:inequality}
    \sum_{t = 1}^{NL} v_{2,t}v_{3,t} \leq \sum_{t = 1}^{NL} v_{2,\sigma(t)}v_{3,t},
\end{equation}
where $v_{(\cdot),t}$ denotes the $t$-th element of vector $\mathbf{v}_{(\cdot)}$ and $\mathbf{v}_{2,\sigma(t)}$ is any permutation of the elements of $\mathbf{v}_{2}$. Since $\mathbf{v}_{2,\sigma(t)}$ is any permutation of $\mathbf{v}_{2}$, which also includes $\mathbf{v}_{1}$, we have proved that OST achieves the minimum transmission delay time for the assumed power distribution, which was optimal for $\mathbf{v}_{1}$. Thus, OST always results in better minimization of \eqref{eq:OptProblemConvex}. 

\bibliographystyle{IEEEtran}
\bibliography{HybridSemShannon}

\end{document}